\documentclass[twocolumn,aps,prb,nofootinbib]{revtex4-2}
\usepackage{graphicx}
\usepackage{times}
\usepackage{bm}
\usepackage[linktocpage=true,colorlinks=true,urlcolor=blue,linkcolor=red,citecolor = blue]{hyperref}
\usepackage{pgfplots}
\usepackage{amsfonts}
\usepackage{amsmath}
\usepackage{amssymb}
\usepackage{xcolor}
\usepackage{braket}
\usepackage{relsize}
\usetikzlibrary{arrows}
\usepackage{enumerate}
\usepackage{pgfplots}
\usepackage{newfloat}
\usepackage{siunitx}
\usepackage{physics}
\usepackage{mathtools}
\usepackage{soul}

\newcommand{\printfnsymbol}[1]{%
  \textsuperscript{\@fnsymbol{#1}}%
}

\DeclareFloatingEnvironment[name={Supplementary Figure}]{suppfigure}

\graphicspath{ {Figures/} }

\begin{document}

%\title{Realizing Majorana zero modes in magnetic field-free InAs-Al nanowires with fewer growth constraints}
\title{Realizing Majorana Kramers pairs in two-channel InAs-Al nanowires with highly misaligned electric fields}
\author{Benjamin D. Woods}
\author{Mark Friesen}
\affiliation{Department of Physics, University of Wisconsin-Madison, Madison, WI 53706, USA}

\begin{abstract}
Common proposals for realizing topological superconductivity and Majorana zero modes in semiconductor-superconductor hybrids require large magnetic fields, which paradoxically suppress the superconducting gap of the parent superconductor. Although two-channel schemes have been proposed as a way to eliminate magnetic fields, geometric constraints make their implementation challenging, since the channels should be immersed in nearly antiparallel electric fields. Here, we propose an experimentally favorable scheme for realizing field-free topological superconductivity, in two-channel InAs-Al nanowires, that overcomes such growth constraints. Crucially, we show that antiparallel fields are not required, if the channels are energetically detuned. We compute topological phase diagrams for realistically modeled nanowires, finding a broad range of parameters that could potentially harbor Majorana zero modes. This work, therefore, solves a major technical challenge and opens the door to near-term experiments.
\end{abstract}

\maketitle

\section{Introduction}
Majorana zero modes (MZMs) are zero-energy modes localized at the ends of topological superconductors and a potential building block for topological qubits~\cite{Bravyi2002,Nayak2008,Alicea2012,Lahtinen2017}. Over the past decade, there has been considerable progress in realizing MZMs in semiconductor-superconductor hybrids~\cite{Mourik2012,Deng2012,Chen2017,Suominen2017,Lutchyn2018,Zhang2021,Microsoft2022,Sarma2023}, primarily in the context of the Lutchyn-Oreg model~\cite{Lutchyn2010,Oreg2010} which requires applying a large magnetic field to the hybrid system. %is immersed in a large magnetic field. 
The application of magnetic fields is worrisome, however, since it suppresses the superconductivity~\cite{Chandrasekhar1962,Clogston1962}. This has motivated research into topological superconductivity in the absence of magnetic fields, so-called time-reversal invariant topological superconductivity (TRITSC)~\cite{Haim2019,Nakosai2012,Volpez2018,Volpez2019,Deng2012b,Zhang2013,Deng2013,Keselman2013,Gaidamauskas2014,Schrade2017,Thakurathi2018,Kotetes2015,Liu2014,Haim2014,Klinovaja2014,Haim2016,Ebisu2016,Parhizgar2017,Oshima2022}, where the MZMs on either end of a wire occur in pairs called Majorana Kramers pairs (MKP), due to the preservation of time-reversal symmetry. 
Of particular interest are proposals that rely on two channels with approximately opposite spin-orbit coupling vectors, such as Rashba bilayers~\cite{Nakosai2012,Volpez2018,Volpez2019} and two-channel nanowires~\cite{Keselman2013,Gaidamauskas2014,Schrade2017,Thakurathi2018,Kotetes2015}, where the system is essentially composed of two copies of the Lutchyn-Oreg model with the spin degree of freedom replaced by the channel degree of freedom. 
%In addition, they require a $\pi$ phase difference between the superconductive pairing in the two channels or interchannel pairing. 

\begin{figure}[t]
\begin{center}
\includegraphics[width=0.48\textwidth]{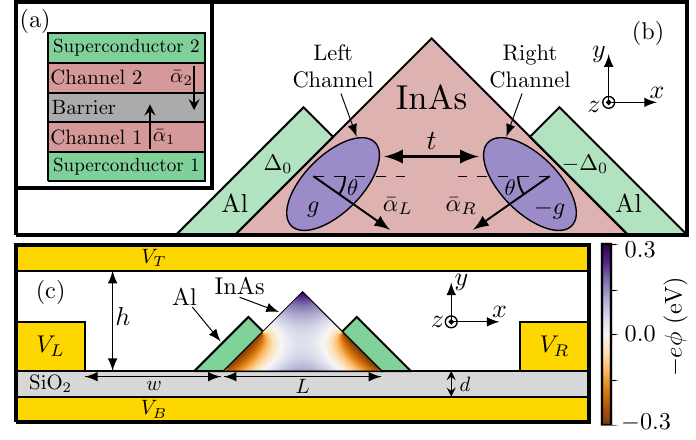}
\end{center}
\vspace{-0.5cm}
\caption{%Schematic of proposed device for the realization of TRITSC. An InAs nanowire (red) is coupled to two Al superconductors (green) whose order parameters differ in phase by $\pi$. Various gates (gold) control the electrostatic environment. Voltages on each gate are in parentheses. A boundary condition $\phi = W$ at the InAs-Al interfaces accounts for the work function difference between the materials. An example potential calculated for $V_{T} = -5.5~\text{V}$, $V_{BG} = -3~\text{V}$, $V_L = V_R = 0$, and $W = 0.3~\text{V}$ is shown on the front cross section of the nanowire, illustrating the importance of the work function difference in forming a channel near each of the InAs-Al interfaces. %Geometric parameters are given in the main text.
\textit{Structures for realizing TRITSC}. (a) Schematic cross section of a proposed Rashba bilayer system, where two superconductors are coupled to two tunnel-coupled channels, with opposite spin-orbit vectors: $\bar{\boldsymbol{\alpha}}_2 = -\bar{\boldsymbol{\alpha}}_1$. %separated by a tunnel barrier (gray). %due to the channels being immersed in opposing electric field.
Here, the spin-orbit vectors align to the local electric field produced by charge inside of the channels.
(b) Schematic cross section of the InAs-Al hybrid nanowire system proposed here, where two channels (blue ovals) with highly misaligned spin-orbit vectors, $\bar{\boldsymbol{\alpha}}_L$ and $\bar{\boldsymbol{\alpha}}_R$, form near the InAs-Al interfaces. Various parameters appearing in the low-energy effective Hamiltonian, Eq.~(\ref{HBdGEff}), are labeled, including the tunnel coupling $t$, energy detuning $g$, spin-orbit misalignment angle $\theta$, and superconductive pairing $\pm \Delta_0$. 
(c) Zoomed out view (not drawn to scale) of the system in (b), showing the gates (gold) used to control the electrostatic environment. The potential $-e\phi$ calculated for the gate voltages $(V_T, V_B, V_L, V_R) = (-5.5, -3, 0, 0)~\text{V}$ and the InAs-Al interface boundary condition $\phi = 0.3~\text{V}$ is plotted in the cross section of the nanowire.}
\label{FIG1}
\vspace{-1mm}
\end{figure}

The planar geometry of Rashba bilayers naturally provides antiparallel spin-orbit vectors in the two channels, as illustrated in Fig.~\ref{FIG1}(a), because the spin-orbit vectors align with the local electric field. %\red{produced by charge inside of the channels}.
Creating such structures in semiconductor-superconductor hybrids is challenging, however, because it requires growing a high-quality semiconducting crystal layer between two superconductors, which 
%to the best of our knowledge 
has yet to be demonstrated. 
In contrast, epitaxial growth of superconductors on semiconductor nanowires has already been accomplished~\cite{Krogstrup2015}. 
In this case, however, it is difficult to engineer antiparallel spin-orbit vectors in the two channels.
This is a serious problem because small misalignments of the spin-orbit vectors are expected to collapse the topological gap~\cite{Keselman2013}, by effectively coupling the two copies of the Lutchyn-Oreg model present in the wire.
% Nonetheless, parallel spin-orbit vectors have been assumed in most works on the subject~\cite{Keselman2013,Gaidamauskas2014,Schrade2017,Thakurathi2018,Ebisu2016}.
%Nevertheless, this is typically assumed in analysis~\cite{Keselman2013,Gaidamauskas2014,Schrade2017,Thakurathi2018,Ebisu2016}, representing a major assumption. Indeed, Ref.~\cite{Keselman2013} showed that a small misalignment of the spin-orbit vectors can destroy the topological gap and MZMs by essentially coupling the two copies of the Lutchyn-Oreg model present in the two-channel system.
%Nonetheless, parallel spin-orbit vectors are typically assumed in theoretical analysis~\cite{Keselman2013,Gaidamauskas2014,Schrade2017,Thakurathi2018,Ebisu2016}.  

%%%%%%%%%%%%%%%%%%%%%%%%%%%%%%%%
%%%%%%%%%%%%%%%%%%%%%%%%%%%%
%%%%%%%%%%%%%%%%%%%%%%	
%%%%%%%%%%%%%%%%%%%%%%%%%%%%%%%%

In this work, we turn the problem on its head by showing that TRITSC can be realized in experimentally feasible~\cite{Vaitiekenas2018,Lee2019} two-channel hybrid nanowires \textit{where the spin-orbit vectors of the two channels are highly misaligned}, as depicted in Fig.~\ref{FIG1}(b). %, well beyond what is found possible in Ref.~\cite{Keselman2013}. 
The key to our proposal is to introduce an energy detuning between the two channels, on order of the superconducting gap. 
%Essentially, the detuning reduces the coupling between the bands of the two copies of the Lutchyn-Oreg model present in the two-channel model by separating them in momentum space. 
Essentially, the detuning separates the two copies of the Lutchyn-Oreg model in momentum space, which significantly reduces the coupling induced by the spin-orbit vector misalignment, allowing the topological phase to persist.
Since the success of our proposal hinges on the degree of spin-orbit vector misalignment and other device parameters, we consider a more rigorous device model than the Hamiltonian-level models typically used in the literature, incorporating both geometric and electrostatic details. This ensures that our conclusions do not result from inaccurate estimates of the model parameters.
%Since the prospect of realizing TRITSC hinges on the degree of spin-orbit vector misalignment and other numerical parameters, we use a model that goes beyond the effective-model level typically found in the literature that incorporates the geometric details of the hybrid structure and the relevant details of the electrostatic environment, including band bending near the semiconductor-superconductor interfaces~\cite{Schuwalow2019}.
%Since geometrical details are at the heart of the problem, we adopt theoretical methods that go beyond the effective model level typically found in the literature by including the the geometric details of the hybrid structure and the relevant details of the electrostatic environment, including gate geometries and band bending near the InAs-Al interfaces~\cite{Schuwalow2019}.
%This more precise treatment ensures that our conclusions do not result from insufficient accuracy.
%This more precise treatment ensures that our conclusions do not result from unreasonable estimates of effective parameters.

\section{Model} \label{Model}
We consider the system shown in Fig.~\ref{FIG1}(c) in which a semiconducting InAs nanowire with a triangular cross section is coupled to two Al superconductors that cover half of the upper facets. % of the nanowire. 
The system is translationally invariant along the $\hat z$ axis. 
Similar to previous  TRITSC proposals~\cite{Keselman2013,Gaidamauskas2014,Schrade2017,Thakurathi2018}, %a phase difference of $\pi$ between the superconducting order parameters is imposed using a flux loop~\cite{Fornieri2019}. 
an external flux loop~\cite{Fornieri2019} imposes an order parameter phase difference of $\pi$ between the two superconductors. 
In addition, gates surround the nanowire to control the electrostatics. %environment. %Note that the exact gate geometry details are unimportant.
The system %is assumed to be translation invariant along the z-axis and 
is modeled by a Bogoliubov-de-Gennes (BdG) Hamiltonian of the canonical form~\cite{Zhu2016,Chiu2016},
\begin{equation}
    H_\text{BdG}(k_z)=
    \begin{pmatrix}
    H_N(k_z) & \widetilde{\Delta} \\
    -\widetilde{\Delta}^* & -H_N^*(-k_z)
    \end{pmatrix}, \label{HBdG}
\end{equation}
where $H_N$ and $\widetilde{\Delta}$ are the normal and superconductive pairing components, respectively, $k_z$ is the momentum along the $\hat z$ axis, and the first and second columns act upon particle and hole degrees of freedom, respectively. The normal Hamiltonian takes the form $H_N = H_0 + H_{\text{SO}}$, where %$H_\text{SO}$ accounts for spin-orbit coupling and 
$H_0$ is an effective mass Hamiltonian given by
\begin{equation}
    %\begin{split}
    H_0(k_z) = %\left(
    \frac{\hbar^2}{2m^*} \left(-\partial_x^2 - \partial_y^2+ k_z^2\right)
    - e\phi(x,y). \label{H0}
    %\right) \sigma_0,% \\ %&+\Big(\alpha_x(\mathbf{r}_\perp)\sigma_y + \alpha_y(\mathbf{r}_\perp)\sigma_x\Big)k_z
    %\end{split},
\end{equation}
Here, $m^*$ is the effective mass %, $e$ is the elementary charge, 
and $\phi$ is the electrostatic potential that satisfies Poisson's equation,
%\begin{equation}
   	 $\nabla \cdot \left[\epsilon(\mathbf{r}) \nabla\phi(\mathbf{r})\right] =
    	  -\rho(\mathbf{r})$, % \label{Pois}
%\end{equation}
where $\epsilon$ is a material-dependent dielectric constant and $\rho$ is the free-charge density that is calculated self-consistently~\cite{Woods2020a}. 
The potential $\phi$ satisfies Dirichlet boundary conditions on the gates shown in Fig.~\ref{FIG1}(c). 
The surfaces of the Al superconductors are also set to $\phi=0$, except for the InAs-Al interfaces, which are set to $\phi= 0.3$~V to account for the band offset between the two materials~\cite{Vuik2016,Antipov2018,Mikkelsen2018,Woods2018}. %An example calculation of the potential is shown in Fig.~\ref{FIG1}. The importance of the work function difference is clearly illustrated with the electrons being strongly attracted towards the InAs-Al interfaces.
The spin-orbit term $H_\text{SO}$ is given by
\begin{equation}
    H_\text{SO}(k_z) = 
    \bigl[\widetilde{\alpha}_x(x,y)\sigma_y - \widetilde{\alpha}_y(x,y)\sigma_x\bigr]k_z, \label{HSO}
\end{equation}
where $\widetilde{\alpha}_j$ are Rashba fields and $\sigma_j$ are Pauli spin matrices. We neglect Dresselhaus and transverse-Rashba spin-orbit coupling contributions because they are weak, as explained in Appendix \ref{Dressel}
%the Supplementary Materials~\cite{SM}. %(see the Supplementary Material where we quantify their effect).
Importantly, the Rashba fields are position-dependent and related to the potential by %the relation %$\alpha_j(x,y) = e C \partial_j \phi(x,y)$
$\boldsymbol{\widetilde{\alpha}}(x,y) = e C \nabla \phi(x,y)$~\cite{Wojcik2018,Escribano2020}, where $C$ is a material-dependent constant. 
%Starting from an 8-band Kane Hamiltonian, perturbation theory yields $\alpha_j(x,y) = e C_\text{SO}~ \partial_j \phi(x,y)$, where $C_\text{SO}$ is a constant whose form is given in Ref.~\cite{Wojcik2018}.
%\begin{equation}
    %\alpha_j(x,y) = \frac{-eP^2}{3}\left( 
   % \frac{1}{E_o^2} - \frac{1}{\left(E_o + E_{SO}\right)^2}
    %\right)
    %\partial_j \phi(x,y), \label{Alpha}
%\end{equation}
%where $E_o$, $E_{SO}$, and $P$ are the fundamental gap, spin split-off gap, and conduction-valence band coupling strength, respectively, of the Kane model.
%It is useful to define the spin-orbit vector for any wave function $\psi_m$ as $\bar{\boldsymbol{\alpha}}_{m} = \int |\psi_m(x,y)|^2 \boldsymbol{\alpha}(x,y) \,dx dy$. %, i.e. the diagonal matrix element of the spin-orbit field $\boldsymbol{\alpha}(x,y)$.
Finally, the superconductive pairing is treated as an induced pairing within the InAs, with $\widetilde{\Delta} = i \Delta(x,y) \sigma_y$. %, where $\Delta(x,y) = \Delta_0 sgn(x)$ and $\Delta_0 \in \mathbb{R}$.
(See details below.)
%where $\Delta$ is the position dependent induced pairing within the semiconductor %. For simplicity, we assume the induced pairing given by
%and given by $\Delta(x,y) = \Delta_0 sgn(x)$,
%\begin{equation}
%    \Delta(x,y) =
%    \begin{cases}
%        -\Delta_0, & x < 0 \\
%        +\Delta_0, & x > 0
%    \end{cases},
%\end{equation}
%where $\Delta_0 \in \mathbb{R}$. %This pairing function is meant to model the $\pi$-phase difference between the two superconductors. Note that while treating the superconductor explicitly within the Hamiltonian is known to renormalize the SM effective parameters~\cite{Stanescu2017a}, it greatly increases computational costs and is not expected to change the qualitative nature of our results. 

%\section{two-channel Effective Model}

\subsection{Two-Channel Effective Hamiltonian} 
We derive the low-energy effective Hamiltonian of the system by considering the electrostatic potential shown in Fig.~\ref{FIG1}(c), for typical gate voltages. 
Here, electrons are attracted to the InAs-Al interfaces, %. Note that this primarily arises from the band offset $W$ between the materials. 
primarily because of the band offset between the materials, and we observe channels forming near each InAs-Al interface, as illustrated in Fig.~\ref{FIG1}(b).
%This attraction causes the formation of a channel near each InAs-Al interface as illustrated in Fig.~\ref{FIG1}(b). 
%Focusing on the low subband-occupancy regime, 
We then project Eq.~(\ref{HBdG}) onto the two lowest-energy orbital subbands, $\varphi_1$ and $\varphi_2$, defined by $H_0(0) \varphi_n(x,y) = \varepsilon_n \varphi_n(x,y)$, where $\varphi_1$ and $\varphi_2$ can be expressed as superpositions of the left and right channel wave functions, $\chi_L$ and $\chi_R$, depicted in Fig.~\ref{FIG1}(b): $\varphi_n = a_n \chi_L + b_n \chi_R$ for $n = 1,2$. In this two-channel basis, as shown in Appendix \ref{appA}, the effective BdG Hamiltonian becomes
\begin{equation}
    \begin{split}
    H_{\text{eff}}(k_z) =&
        \left[
        \varepsilon(k_z)
         \lambda_0  
        + t \lambda_x 
        + g \lambda_z 
        \right]\sigma_0\tau_z 
        - \Delta_0 \sigma_y \lambda_z \tau_y
        \\
        &+ \alpha k_z \left(\cos\theta \sigma_y \lambda_z \tau_z 
        +  \sin\theta \sigma_x \lambda_0 \tau_0\right) ,
    \end{split} \label{HBdGEff}
\end{equation}
where $\sigma_j$, $\lambda_j$, and $\tau_j$ are Pauli matrices acting on spin, channel, and particle-hole space, respectively, and $\varepsilon(k_z) = \hbar^2 k_z^2/(2m^*) - \mu$ is the bare dispersion. 
By defining $h_{ij} = \mel{\chi_i}{H_0(0)}{\chi_j}$, we can express the chemical potential as $\mu = (h_{LL}\!+\!h_{RR})/2$, the interchannel tunnel coupling as $t = h_{RL}$, and the channel energy detuning as $g= (h_{LL}\!-\!h_{RR})/2$. We note that these parameters all depend on the local electrostatics.
For example, $t$ characterizes the localization of the channels near the InAs-Al interfaces, while $g$ is largely determined by the voltage bias $V_L-V_R$ between the side gates, shown in Fig.~\ref{FIG1}(c).
%%%%%%%%%%%%%%%%%%%%%%%%%%%%%%%%
%%%%%%%%%%%%%%%%%%%%%%%%%%%%
%\begin{figure}[t]
%\begin{center}
%\includegraphics[width=0.48\textwidth]{Figures/FigSchematic.pdf}
%\end{center}
%\vspace{-0.5cm}
%\caption{Schematic of the two-channel effective model. Two channels (blue ovals) form near the InAs-Al interfaces due to a work function difference between the materials. There exists interchannel tunnel coupling $t$ from wave function overlap and an energy detuning $g$ between the two channels from a voltage bias between the side gates shown in Fig.~\ref{FIG1}. Each channel has a spin-orbit vector $\bar{\boldsymbol{\alpha}}_{m}$ due to a local electric field, where the misalignment of these vectors is given by the angle $2\theta$. Finally, the channels have superconductive pairing $\pm \Delta_0$ due to a $\pi$ phase difference between the superconductors.}
%label{FIGSchematic}
%\vspace{-1mm}
%\end{figure}
%%%%%%%%%%%%%%%%%%%%%%	
%%%%%%%%%%%%%%%%%%%%%%%%%%%%%%%%
To understand the parameters $\alpha$ and $\theta$ in Eq.~(\ref{HBdGEff}), it is helpful to define the spin-orbit vector of channel $i$ as $\bar{\boldsymbol{\alpha}}_{i} = \int |\chi_i(x,y)|^2 \boldsymbol{\widetilde{\alpha}}(x,y) \,dx dy$, which averages the electric field over the channel wave function. 
The parameters $\alpha$ and $\theta$ appearing in Eq.~(\ref{HBdGEff}) represent the magnitude and direction of the spin-orbit vectors, as illustrated in Fig.~\ref{FIG1}(b). Crucially, the sample geometry ensures that $\bar{\boldsymbol{\alpha}}_L$ and $\bar{\boldsymbol{\alpha}}_R$ are generically misaligned. Additionally, the symmetry of the nanowire cross section imparts approximate mirror symmetry to the two spin-orbit vectors, such that $\bar{\boldsymbol{\alpha}}_{L}\cdot \hat x\approx -\bar{\boldsymbol{\alpha}}_{R}\cdot \hat x$ and $\bar{\boldsymbol{\alpha}}_{L}\cdot \hat y\approx \bar{\boldsymbol{\alpha}}_{R}\cdot \hat y$.
This is reflected %in the second line of Eq.~(\ref{HBdGEff}) by the $\lambda_j$ factors in the $\cos\theta\sigma_y\lambda_z$ and $\sin\theta\sigma_x\lambda_0$ terms.
in the $\lambda_z$ and $\lambda_0$ factors in the second line of Eq.~(\ref{HBdGEff}). 
The angle $\theta$ characterizes the misalignment between the spin-orbit vectors, and how far the system is from the \textit{ideal limit} ($\theta = 0$), 
%of two channels with exactly \textit{opposite} spin-orbit vectors. Note that the \textit{ideal limit} ($\theta = 0$) is 
typically assumed in the literature~\cite{Nakosai2012,Volpez2018,Volpez2019,Gaidamauskas2014,Schrade2017,Thakurathi2018,Kotetes2015}. 
In a real device, the spin-orbit vectors will not be perfectly mirror symmetric, due to the voltage bias $V_L-V_R$ and device imperfections. 
We neglect such symmetry-breaking terms here for simplicity, and because they do not significantly affect the physics, as shown in Appendix \ref{appA}.
%~\cite{SM}.
%Finally, the superconductive pairing is opposite for the two channels due to the imposed $\pi$ phase difference between the two Al superconductors as shown in Fig.~\ref{FIG1} (b). % and indicated by the $\lambda_z$ factor in the $\Delta_0$ term of Eq.~(\ref{HBdGEff}).
Finally, we note that the superconductive pairing for channel $i$ could also be computed as $\Delta_i = \int |\chi_i(x,y)|^2 \Delta(x,y) \,dx dy$, in principle. %where we can simply compute the integrals if $\Delta(x,y)$ is specified. 
%Obtaining meaningful results from such calculations, however, would require explicit inclusion of the Al regions within the Hamiltonian~\cite{Stanescu2017a,Antipov2018,Stanescu2022}, where $\Delta(x,y) \neq 0$ only within the Al regions. Since we are treating superconductivity as an induced pairing within the InAs, it is not obvious what form $\Delta(x,y)$ should take. We avoid these complications by instead imposing $\Delta_L, \Delta_R = \Delta_0, -\Delta_0$. Importantly, this choice still captures the essential physics of the channels having induced superconductivity with a $\pi$ phase difference. %, while avoiding the complications of explicitly including the Al in the Hamiltonian.
However, a rigorous calculation of $\Delta(x,y)$ involves complicating factors, such as details about disorder and the InAs-Al interface~\cite{Stanescu2022}, which are beyond the scope of this work.
%It is not obvious, however, what functional form $\Delta(x,y)$ should take and would require more complicated model~\cite{Stanescu2017a} beyond the scope of this work. 
Here, we simply define ($\Delta_L, \Delta_R) = (\Delta_0, -\Delta_0$), which captures the $\pi$ phase difference between the superconducting order parameters, and we adopt a typical value of $\Delta_0 = 0.2~\text{meV}$ for such InAs-Al hybrids~\cite{Lutchyn2018}. %Importantly, this simplification still captures the essential physics of the channels having induced superconductivity with a $\pi$ phase difference.

\section{Topological Phase Diagrams}
Before focusing on the device shown in Fig.~\ref{FIG1}(c), we first explore how the channel degree of freedom enables the realization of TRITSC. Additionally, we show how misalignment of the spin-orbit vector ($\theta \neq 0$) is detrimental to the topological phase and how introducing a channel detuning ($g \neq 0$) can alleviate this issue. 
For now, the parameters $\mu$, $t$, $g$, and $\theta$ in Eq.~(\ref{HBdGEff}) are treated as free parameters, to elucidate the physics of the effective model. However, they are calculated more realistically, later in the paper. Other parameters used here are $m^* = 0.026 m_0$ and $\alpha =20~\text{meV}\cdot\text{nm}$~\cite{Lutchyn2018}. 

%%%%%%%%%%%%%%%%%%%%%%%%%%%%%%%%
%%%%%%%%%%%%%%%%%%%%%%%%%%%%
\begin{figure}[t]
\begin{center}
\includegraphics[width=0.48\textwidth]{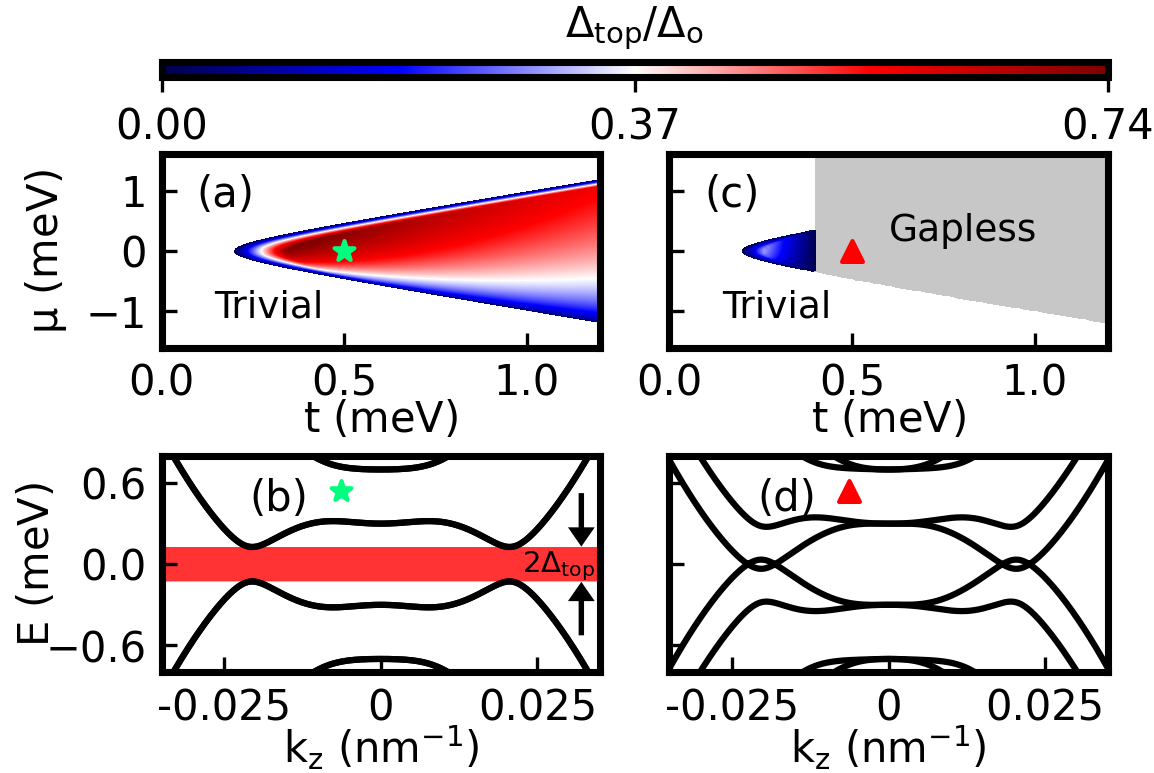}
\end{center}
\vspace{-0.5cm}
\caption{%a) Topological phase diagram in the limit of parallel spin-orbit vectors ($\theta = 0$) and zero channel detuning ($g = 0$). The colorful region indicates the topologically non-trivial phase with the color map indicating the size of the topological gap $\Delta_{\text{top}}$. b) Bulk spectrum for parameters indicated by the green star in (a). Note that all bands are double degenerate. c) Phase diagram for the same system as (a) except $\theta = 30^\circ$. The green region indicates a gapless phase. d) Bulk spectrum for parameters indicated by the red triangle in (c). Notice that the spectrum is gapless. Parameters for all panels are $m^* = 0.026 m_o$, $\Delta_0 = 0.2~\text{meV}$, and $\alpha =20~\text{meV}\cdot\text{nm}$.
\textit{Topological phase diagrams in the absence of channel detuning}. (a) Topological phase diagram for antiparallel spin-orbit vectors ($\theta = 0$) and no channel detuning ($g = 0$). The colormap shows the computed topological gap $\Delta_{\text{top}}$ for the topological phase. 
(b) Bulk energy spectrum for parameters corresponding to the green star in (a). All bands are two-fold degenerate because Eq.~(\ref{HeffTrans}) decomposes into two identical spin blocks when $\theta \!= g\!= 0$. The shaded energy gap corresponds to $2\Delta_{\text{top}}$. 
(c) Phase diagram for the same system as (a), but with spin-orbit misalignment $\theta = 30^\circ$, yielding a large gapless region (gray). 
(d) A typical gapless bulk spectrum, for parameters corresponding to the red triangle in (c).} %Parameters for all panels are $m^* = 0.026 m_o$, $\Delta_0 = 0.2~\text{meV}$, and $\alpha =20~\text{meV}\cdot\text{nm}$.}
\label{FIG2}
\vspace{-1mm}
\end{figure}
%%%%%%%%%%%%%%%%%%%%%%	
%%%%%%%%%%%%%%%%%%%%%%%%%%%%%%%%

To begin, it is helpful to change the basis of the effective Hamiltonian, Eq.~(\ref{HBdGEff}), as explained in Appendix \ref{TransApp}, yielding 
%the unitary matrix $U = (1 + i[\sigma_x + \sigma_y + \sigma_z])(1 - i\sigma_z\lambda_x)(1 + i\lambda_x\tau_z)/4$. The transformed Hamiltonian $H_{\text{eff}}^\prime(k_z) = U^\dagger H_{\text{eff}}(k_z) U$ is found to be 
\begin{equation}
    \begin{split}
    H_{\text{eff}}^\prime(k_z) =&
    [
    \varepsilon(k_z) \lambda_0 \tau_z  
        + t \lambda_x \tau_z 
        - \Delta_0 \lambda_y \tau_y \\
        &+ \alpha k_z \cos\theta \lambda_z \tau_0
    ] \sigma_0  + g \sigma_z \lambda_z \tau_0  \\
    &+ \alpha k_z \sin\theta \sigma_x \lambda_x \tau_0 .
    \end{split} \label{HeffTrans}
\end{equation}
%where $H_{\text{eff}}^\prime(k_z) = U^\dagger H_{\text{eff}}(k_z) U$ and $U$ is a unitary matrix discussed in the Supplementary Material.
We first consider the limit of antiparallel spin-orbit vectors ($\theta = 0$), as typically assumed in the literature~\cite{Gaidamauskas2014,Schrade2017,Thakurathi2018,Kotetes2015}, with no channel detuning ($g = 0$). 
Equation~(\ref{HeffTrans}) then decomposes into two identical spin blocks. 
[This is the reason for the basis change in Eq.~(\ref{HeffTrans}).]
As noted previously~\cite{Keselman2013}, each of the spin blocks corresponds to an independent Lutchyn-Oreg model~\cite{Lutchyn2010,Oreg2010} of a superconducting nanowire with spin-orbit coupling in an external magnetic field, except that the spin degrees of freedom are now replaced by channels $\sigma_j \rightarrow \lambda_j$, and the Zeeman energy is replaced by the interchannel tunnel coupling  $t$.
%The Lutchyn-Oreg model is known to undergo a topological phase transition with the closing of the bulk gap at $k_z = 0$ when $t^2 = \mu^2 + \Delta_0^2$. For $t^2 > \mu^2 + \Delta_0^2$, the bulk gap reemerges (now called the topological gap $\Delta_\text{top}$), and the system is in a topological phase that contains a MZM at each end of the system.
In analogy with the Lutchyn-Oreg model, the system transitions from a trivial phase without MZMs to a topological phase with MZMs when $t^2 > \mu^2 + \Delta_0^2$. In contrast to the Lutchyn-Oreg model, however, \textit{two MZMs (corresponding to one MKP) appear at each end of the nanowire}, due to the presence of dual spin blocks in the TRITSC case. 
(Note that the degeneracy of the two pairs of MZMs at zero energy is protected by time-reversal symmetry~\cite{Haim2019}.)
%Note that since this degeneracy is protected by time-reversal symmetry, these Majorana zero modes are often called Majorana Kramers pairs.
An example phase diagram is shown in Fig.~\ref{FIG2}(a) for the case of $\theta\!= g \!= 0$. 
(See Appendix~\ref{PDCMapp} for a discussion of our method for calculating phase diagrams, which involves determining whether an MKP exists at the end of a semi-infinite wire.)
Like other MZM schemes, several device parameters (including the chemical potential) must be tuned simultaneously, to enter the topological phase. Indeed, the range of the chemical potentials yielding a topological phase in Fig.~\ref{FIG2}(a) is the same range found for Majorana nanowires with broken time-reversal symmetry \cite{Lutchyn2010,Oreg2010}.
%In addition, an example of the bulk spectrum in the topological phase is shown in Fig.~\ref{FIG2}(b). %Since we are considering $\theta \!= g \!= 0$ here, the Hamiltonian is composed of two equivalent blocks, which yields a spectrum that is double degenerate everywhere.
%Since the Hamiltonian for $\theta \!= g \!= 0$ is composed of two identical spin blocks, 
In Fig.~\ref{FIG2}(a), the colormap indicates the size of the topological gap $\Delta_{\text{top}}$, defined by the red region in Fig.~\ref{FIG2}(b), where we plot a typical bulk energy spectrum. 
We note that this spectrum is everywhere two-fold degenerate, due to the identical spin blocks.

%%%%%%%%%%%%%%%%%%%%%%%%%%%%%%%%
%%%%%%%%%%%%%%%%%%%%%%%%%%%%
\begin{figure}[t]
\begin{center}
\includegraphics[width=0.48\textwidth]{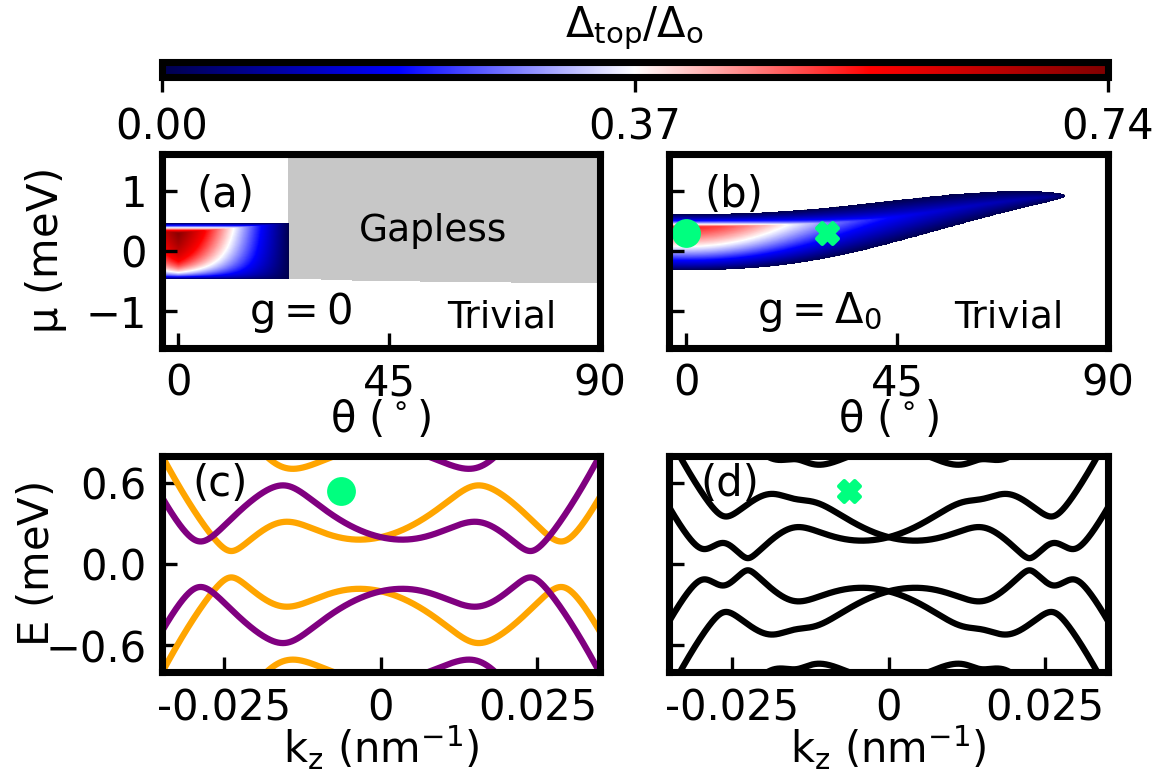}
\end{center}
\vspace{-0.5cm}
\caption{\textit{Effect of channel detuning $g$ on topological phase diagrams}. (a), (b) Topological phase diagrams, as function of the chemical potential $\mu$ and spin-orbit misalignment $\theta$, for $t = 0.5~\text{meV}$, and (a) $g = 0$, or (b) $g = \Delta_0$.
Increasing $\theta$ in (a) suppresses the topological gap $\Delta_\text{top}$, yielding a gapless phase (gray) when $\theta \gtrsim 22^\circ$.
Introducing the channel detuning $g$ in (b) removes the gapless phase and extends the topological region out to large $\theta$. 
%spin-orbit misalignment angles $\theta$.
(c), (d) Bulk spectra for parameters corresponding to the green circle ($\theta = 0$) and green cross ($\theta = 30^\circ$) in (b). %, where $\mu = 0.3~\text{meV}$. Comparing this to Fig.~\ref{FIG2} (b), we see that including a channel detuning $g$ breaks the spin degeneracy away from $k_z = 0$ with spin $\uparrow$ and $\downarrow$ being orange and purple lines, respectively.
In (c), %$g \neq 0$
the channel detuning $g$ lifts the spin degeneracy when $k_z \neq 0$, as shown with orange (purple) for the spin $\uparrow$ ($\downarrow$) bands.
%Comparing (c) to Fig.~\ref{FIG2} (b), we see that the energy detuning $g$ breaks the spin degeneracy away from $k_z = 0$ with spin $\uparrow$ and $\downarrow$ being orange and purple lines, respectively. 
Lifting the degeneracy reduces the level repulsion between spins, caused %by nonparallel spin-orbit vectors ($\theta \neq 0$), 
by misaligned spin-orbit vectors, and allows the topological phase to survive up to large $\theta$.}
%Parameters for all panels are $m^* = 0.026 m_o$, $\Delta_0 = 0.2~\text{meV}$, and $\alpha =20~\text{meV}\cdot\text{nm}$.}
\label{FIG3}
\vspace{-1mm}
\end{figure}
%%%%%%%%%%%%%%%%%%%%%%	
%%%%%%%%%%%%%%%%%%%%%%%%%%%%%%%%

Next, we consider the case of misaligned spin-orbit vectors ($\theta \neq 0$), although we still assume $g = 0$. 
%Again, we stress that this geometry is more realistic for the wire geometry shown in Fig.~\ref{FIG1}(c). %as shown in Fig.~\ref{FIGSchematic}. 
%First, we note that the closing and opening of the bulk energy gap responsible for the topological phase transition in the ideal limit occurs at $k=0$. Now since the $\sin(\theta)$ term in Eq.~(\ref{HBdGEff}) vanishes at $k = 0$, we know that making $\sin(\theta) \neq 0$ cannot affect the topological phase transition at $k = 0$. It can impact the topological gap, however, which occurs at $k \neq 0$. 
%While not affecting the topological phase transition which occurs with the closing of the bulk gap at $k_z = 0$, letting $\theta \neq 0$ does affect the topological gap, which occurs at $k_z \neq 0$. Indeed, the $\sin\theta$ term couples the two blocks that were previously decoupled in the ideal limit. This, in turn, breaks the degeneracy of the bulk spectrum and reduces the topological gap. 
The term $\alpha k_z\! \sin \theta \sigma_x \lambda_x \tau_0$ in Eq.~(\ref{HeffTrans}) now plays an important role, coupling the two spin blocks. 
This has a dramatic effect on the phase diagram, as shown in Fig.~\ref{FIG2}(c) for $\theta = 30^\circ$. Here, the topological phase \textit{boundary} is unchanged at small $t$; however, the topological \textit{gap} is significantly reduced, disappearing all together for large enough $t$. Importantly, the closing of the gap destroys the topological phase and accompanying MZMs because a topological phase is only well-defined in the presence of a bulk gap~\cite{Stanescu2017}. This explains why misaligned spin-orbit vectors are considered dangerous for TRITSC~\cite{Keselman2013}. 
%\red{\st{The closing of the gap also destroys the MZMs, which rely on the presence of a topological phase}~\cite{Stanescu2017}.} 
%The closing of the gap also destroys the MZMs because a topological phase and its topological boundary modes require a bulk gap~\cite{Stanescu2017}. 
A typical gapless bulk spectrum is shown in Fig.~\ref{FIG2}(d). 
Here, the coupling between the spin blocks, due to spin-orbit vector misalignment, causes the degenerate bands to split, when $k_z \neq 0$.

The key result of this paper is that the topological gap can remain robust, even for highly misaligned spin-orbit vectors, %if the mirror symmetry about the y axis is broken 
if we introduce an energy detuning $g$ between the two channels. This is demonstrated in Figs.~\ref{FIG3}(a) and \ref{FIG3}(b), where we report topological phase diagrams for different channel detunings, as a function of the spin-orbit misalignment angle $\theta$ and chemical potential $\mu$. 
For the case $g = 0$ [Fig.~\ref{FIG3}(a)], we see that increasing $\theta$ suppresses the topological gap, until the system becomes gapless for $\theta \gtrsim 22^\circ$, as consistent with Fig.~\ref{FIG2}(c). 
In stark contrast for $g = \Delta_0$ [Fig.~\ref{FIG3}(b)], there is no gapless phase, and the topological phase remains intact for a range of chemical potentials that is nearly constant for $\theta$ between 0 and $40^\circ$. In fact, the topological phase extends nearly to
%nearly extending to 
the \textit{worst-case limit} of $\theta = 90^\circ$, where the spin-orbit vectors of the two channels are parallel. 
Detuning is therefore found to be crucial for maintaining a topological phase in TRITSC when $\theta \neq 0$. 

To understand the mechanism providing robustness to the topological gap, we first consider the case of $\theta = 0$ and $g \neq 0$. %, which is accounted for in the effective Hamiltonian in Eq.~(\ref{HeffTrans}) by the term $g \sigma_z \lambda_z \tau_0$. 
Solving for the gap-closing condition, we obtain the topological criterion, $t^2 > (\mu - E_g)^2 + \Delta_0^2$, where $E_g = \hbar^2 g^2/(2m^*\alpha^2)$. Interestingly, we find that the bulk gap does not close at $k_z = 0$, as in the Lutchyn-Oreg model, but rather at $k_z = \mp g/\alpha$. 
More importantly, the topological criterion is the same as the $g = 0$ case, except for an overall shift to larger chemical potentials. Therefore, introducing the channel detuning $g$ by itself does not alter the size of the topological region. 
We then consider the spectrum for $\theta = 0$ and $g = \Delta_0$, in the topological regime [Fig.~\ref{FIG3}(c)]. Comparing this to the $g = 0$ spectrum in Fig.~\ref{FIG2}(b), we see that the channel detuning $g$ breaks the two-fold spin degeneracy when $k_z \neq 0$.
This occurs because the $g \sigma_z\lambda_z \tau_0$ term in Eq.~(\ref{HeffTrans}) renders the two spin blocks inequivalent. 
If we also misalign the spin orbit vectors ($\theta \neq 0$), as shown in Fig.~\ref{FIG3}(d) for the case $\theta = 30^\circ$, we find that the topological gap is reduced, as expected from Fig.~\ref{FIG2}.
However, the topological gap is more robust, because the band minima are separated in $k_z$ space [see Fig.~\ref{FIG3}(c)], effectively reducing the %mixing between the two spin blocks arising from spin-orbit vector misalignment. 
level repulsion between the two spin blocks when $\theta \neq 0$. Finally, we note that, to be effective, $g$ must be comparable to other energies in the problem, which are of order of the induced gap $\Delta_0$.

\section{Full Device Calculations}
We now compute topological phase diagrams for the device shown in Fig.~\ref{FIG1}(c), as well as the parameters $t$, $\theta$, and $\alpha$ appearing in the effective Hamiltonian, Eq.~(\ref{HeffTrans}). 
The results suggest that prospects for realizing TRITSC in such a device are very promising, and validate the conclusions of Figs.~\ref{FIG2} and \ref{FIG3} in a realistic setting. %We stress that, here, the parameters $t$, $g$, $\theta$, and $\alpha$ of the effective Hamiltonian in Eq.~(\ref{HBdGEff}) are \textit{not} free parameters. Rather, they are dictated by the geometry of the nanowire and the electrostatic environment. 
For these calculations, we adopt the geometric parameters $(L,d,h,w) = (65, 50,90, 100)~\text{nm}$ 
%\red{\st{with atomically sharp interfaces}} 
and assume a 45$^\circ$ angle at the base of the nanowire, as consistent with recent experiments~\cite{Lee2019}. 
The triangular cross section is found to be beneficial for TRITSC by providing single-subband occupancy in the conduction band, while maintaining a large interchannel tunnel coupling $t$, due to the large surface-to-volume ratio.
%, and avoiding the occupation of holes in the valence band. %This helps avoid the subband problem that often plauges Majorana nanowire devices~\cite{Chen2019,Woods2019b,Woods2020a,Pan2020b}. 
As before, the band offset at the InAs-Al interfaces is set to $0.3~\text{V}$~\cite{Schuwalow2019}, %The induced pairing is $\Delta_0 = 0.2~\text{meV}$, typical for InAs/Al hybrid nanowires~\cite{Lutchyn2018}. 
and other parameters are given by $C = 1.17~\text{nm\textsuperscript{2}}$ and $(\epsilon_{\text{InAs}}, \epsilon_{\text{Si} \text{O}_2}) = (14.6, 3.9)$~\cite{Winkler2003}. 
Our numerical procedure begins by self-consistently~\cite{Woods2020a} solving the Schr\"odinger-Poisson equations, treating the charge density in the Hartree approximation. Here, the orbital subbands $\varphi_n$ of $H_0$ in Eq.~(\ref{H0}) are solved using finite-element methods~\cite{Woods2020c} and the electrostatic potential is solved with help from the FEniCS software library~\cite{Alnaes2015}. Note that we focus on the very-low-density regime, where self-consistency has only a modest impact on the results. We then use the solutions obtained for $\varphi_n$ to calculate the matrix elements of $H_{\text{SO}}$ and $\widetilde{\Delta}$. Projecting onto just a few ($n \lesssim 10$) low-energy subbands, we finally calculate the local density of states near zero energy, for a semi-infinite wire, as described in Appendices \ref{appA} and \ref{PDCMapp}. This procedure is repeated for a range of system parameters to obtain a topological phase diagram.

\begin{figure}[t]
\begin{center}
\includegraphics[width=0.48\textwidth]{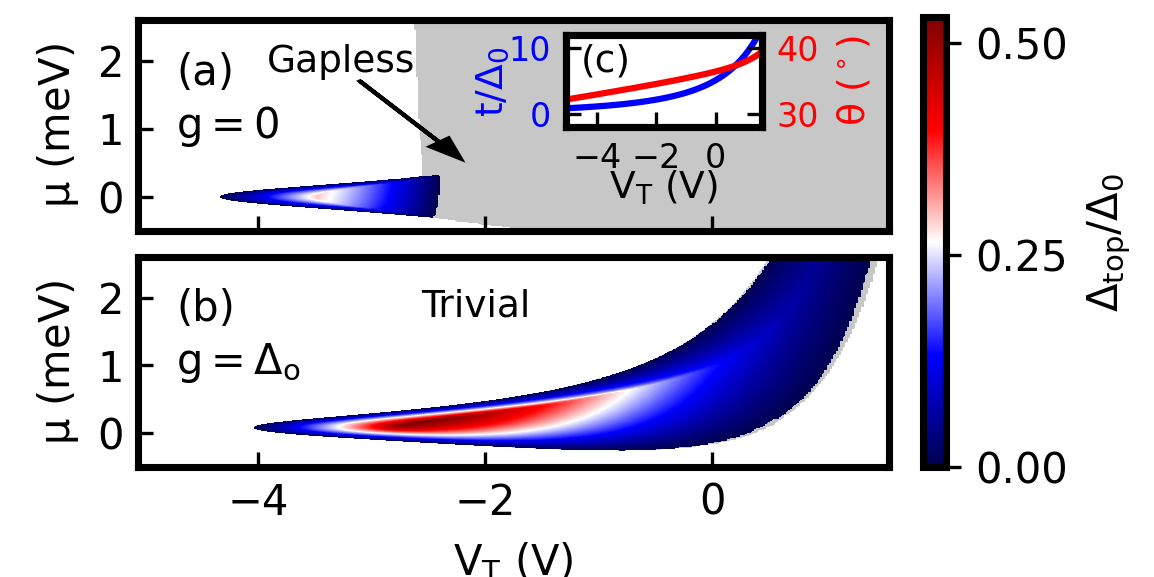}
\end{center}
\vspace{-0.5cm}
\caption{%Topological phase diagrams of the device shown in Fig.~\ref{FIG1} (c) as a function of $V_{T}$ and $\mu$ for systems with (a) $g = 0$ and (b) $g = \Delta_0$. $\mu = 0$ corresponds to the backgate voltage $V_{BG}$ of the dashed line in the 1 occupied subband region of Fig.~\ref{FIG4} (a). Moving along $\mu$ corresponds to moving perpendicular to the dashed line in Fig.~\ref{FIG4} (a). A non-zero detuning $g$ arises by applying a voltage difference between the side gates, $V_L$ and $V_R$. For the case of $g = 0$ in (a), we have a large gapless region and only a small topologically non-trivial region with a small topological gap. Introducing the energy detuning $g = \Delta_0$ in (b) removes the gapless region, increases the size of the topologically non-trivial region, and increases the topological gap.}
\textit{Topological phase diagrams for full device}. 
Phase diagrams are shown for the device in Fig.~\ref{FIG1}(c), as functions of topgate voltage $V_T$ and chemical potential $\mu$, for channel detuning values of (a) $g = 0$, or (b) $g = \Delta_0$. Similar to Figs.~\ref{FIG3}(a) and \ref{FIG3}(b), we find a large gapless region (gray) in (a) for $g = 0$, which is nearly removed in (b) by introducing a channel detuning $g = \Delta_0$. 
Inset: Normalized tunnel coupling $t/\Delta_0$ (blue) and spin-orbit misalignment angle $\theta$ (red), which characterize the effective Hamiltonian in Eq.~(\ref{HBdGEff}), and are numerically extracted as a function of topgate voltage $V_T$, for $\mu = g= 0$. %The increase of $t$ with increasing $V_T $ explains the broadening of the topological phase along the $\mu$ axis with increasing $V_T$.
Note that the width of the topological region, as a function of $\mu$, is largely determined by $t$.
Also note that $\theta \approx 35^\circ$ for all $V_T$, indicating highly misaligned spin-orbit vectors. 
Parameters like $t$ and $\theta$ depend strongly on the device geometry and electrostatics.}
\label{FIG6}
\vspace{-1mm}
\end{figure}
%%%%%%%%%%%%%%%%%%%%%%	
%%%%%%%%%%%%%%%%%%%%%%%%%%%%%%%%

Figures~\ref{FIG6}(a) and \ref{FIG6}(b) compare the topological phase diagrams obtained for the different channel detuning values, (a) $g = 0$ and (b) $g = \Delta_0$. %Here, $\mu = 0$ corresponds to the backgate voltage $V_{BG}$ of the dashed line in the 1 occupied subband region of Fig.~\ref{FIG4} (a). Moving along $\mu$ corresponds to moving perpendicular to the dashed line in Fig.~\ref{FIG4} (a). We achieve a non-zero detuning $g$ by applying a voltage difference between the side gates. 
Here, the backgate and topgate provide full control over the chemical potential $\mu$, where $\mu = 0$ is defined as the Fermi energy lying halfway between the first two subband energies, at $k_z = 0$.
%Here, the backgate voltage $V_{B}$ tunes the chemical potential $\mu$ for any given value of $V_{T}$. %Note that $\mu = 0$ corresponds to having one occupied subband with the subband energies %for $k_z = 0$ 
%being $\varepsilon_n = \pm \sqrt{t^2 + g^2}$ for the two lowest-energy subbands. %, $n = 1,2$. %See the Supplementary Material for a subband occupancy plot. 
%Note that $\mu = 0$ corresponds to the Fermi energy being exactly between the first two subband energies at $k_z = 0$. 
%Also note that the channel detuning $g$ is tuned by adjusting the voltage bias between $V_L$ and $V_R$ in Fig.~\ref{FIG1}(c). 
The phase diagrams can be understood in analogy to Figs.~\ref{FIG2} and \ref{FIG3}. In short, Fig.~\ref{FIG6}(a) shows a large gapless region, due to the absence of channel detuning ($g = 0$). However, this region almost disappears in \ref{FIG6}(b), when we include a channel detuning $g = \Delta_0$ by applying a voltage bias $V_L-V_R$. 
The striking difference between Figs.~\ref{FIG6}(a) and \ref{FIG6}(b) can be attributed to the channel detuning %, when the spin-orbit coupling vectors are misaligned 
in the presence of highly misaligned spin-orbit vectors.
Here, the computed misalignment angle $\theta$ is shown in the inset (red curve), and is found to range from $32^\circ$ to $39^\circ$, remaining close to $45^\circ$ due to the geometry. 
%This is comparable to the angle $\gamma = 45^\circ$ at the base of the nanowire since the electric field is required by Poisson's equation to be normal at the InAs-Al interfaces. 
The topological gap in Fig.~\ref{FIG6}(b) is also large, %reaching a maximum of $\Delta_{\text{top}} \approx 0.5 \Delta_0$, 
indicating a robust topological phase. Moreover, the phase diagram is quite broad with respect to $\mu$, particularly for larger $V_T$ values, due to the exponential dependence of the interchannel tunnel coupling $t$ on $V_T$, as shown in the inset (blue curve). 
%Note that the large tunability of $t$ comes from the malleability of the electrostatic potential as the gates $V_T$ and $V_B$ are changed, effectively altering the confinement of the channels near the InAs-Al interfaces. 
We note that a broad phase diagram is particularly conducive for forming MZMs, since larger fluctuations in $\mu$ can be tolerated along the length of the wire, without admitting localized Andreev bound states~\cite{Sau2013a,Rainis2013,Pan2020,Prada2020,Woods2021a,Zeng2021,Kells2012,Stanescu2019a,Vuik2019}. 
%Finally, other features of the phase diagrams can be understood by studying the effective parameters $t$ (blue line) and $\theta$ (red line) shown in the inset. %, which are numerically extracted for $V_L = V_R = 0$. 
%We see that the tunnel coupling $t$ monotonically increases with $V_T$. %, reaching a maximum value of $t \approx 2~\text{meV}$. 
%This explains why the width of the topological regions in both Fig.~\ref{FIG6}(a) and (b) grow in $\mu$ space with increasing $V_T$ since the topological condition $t^2 > (\mu- E_g)^2 + \Delta_0^2$ must be satisfied. Physically, the increase of $t$ %with increasing $V_T$ 
%occurs due to a decreasing confinement of the channels near the InAs-Al interfaces as we tune the gates.
%In addition, we find the spin-orbit misalignment angle $\theta$ (red line in the inset) ranges from $32^\circ$ to $39^\circ$, representing highly nonparallel spin-orbit vectors. 
%This is comparable to the angle $\gamma = 45^\circ$ at the base of the nanowire since the electric field is required by Poisson's equation to be normal at the InAs-Al interfaces. 
Finally, we obtain a spin-orbit magnitude of $\alpha\approx 25~\text{meV}\cdot\text{nm}$ for the range of voltages shown in Fig.~\ref{FIG6}, a spin-orbit energy of $E_{\text{SO}} = m^* \alpha^2/(2\hbar^2) \approx 0.11~\text{meV}$, and a spin-orbit length of $l_\text{SO} = \hbar^2/(m^* \alpha) \approx 115~\text{nm}$. This large spin-orbit coupling arises mainly from the electric field induced by the band offset at the InAs-Al interfaces.

\section{Conclusion}
Contrary to previous expectations~\cite{Keselman2013}, we have shown how TRITSC may be realized in InAs-Al two-channel nanowires with highly misaligned spin-orbit vectors. The key ingredient in our scheme is an energy detuning $g$ between the two channels. 
By incorporating geometrical and electrostatic details into a realistic device model, we have shown that MZMs can be realized, without magnetic fields, using currently existing InAs-Al technology~\cite{Vaitiekenas2018,Lee2019}. 
%Starting from a realistic model for an InAs-Al hybrid device incorporating the various details of the electrostatics, we have shown that TRITSC %with a topological gap on order half of the original induced gap 
%can be realized in InAs-Al hybrid nanowires hosting two channels with highly nonparallel spin-orbit vectors. This is contrary to what was previously thought possible~\cite{Keselman2013}. The key ingredient that allows such large spin-orbit vector misalignment is an energy detuning $g$ between the two channels. Finally, we stress that our proposal should be within reach of current InAs-Al technology~\cite{Vaitiekenas2018,Lee2019}.

We caution however, that the proposed scheme still requires a high degree of uniformity along the length of the wire to avoid forming low-energy Andreev bound states, similar to other MZM schemes. %~\cite{Kells2012,Sau2013a,Stanescu2019a,Vuik2019,Pan2020,Prada2020,Woods2021a,Zeng2021}
%~\cite{Chen2019}.
Moreover, fluctuations of the interchannel tunnel coupling $t$ and the detuning parameter $g$ could also affect %\st{qubit operations}
the formation of Andreev bound states, in addition to more well-studied fluctuations of the chemical potential $\mu$ %\textcolor{red}{and the confinement potential of the wire, which can include electrostatic fluctuations}
coming from electrostatic nonuniformities \cite{Sau2013a,Rainis2013,Pan2020,Prada2020,Woods2021a,Zeng2021}, including smooth confinement potentials at the edges of the wire
~\cite{Kells2012,Stanescu2019a,Vuik2019}. % Finally, we note that the decoupling of the spin blocks in TRITSC relies upon time-reversal symmetry.
Finally, we note that the decoupling of the MKP in TRITSC relies upon time-reversal symmetry.
Magnetic impurities, such as Overhauser fields from nuclear spins, familiar to the quantum dot community~\cite{Reilly2008}, could therefore be detrimental and will be addressed in future work.

\begin{acknowledgments}
\textit{Acknowledgments} -- We are grateful to M. A. Eriksson and T. D. Stanescu for helpful discussions.
This research was sponsored in part by the Army Research Office (ARO) under Award No.\ W911NF-22-1-0090, and by the National Science Foundation (NSF) through QLCI-HQAN (Award No.\ 2016136). The views, conclusions, and recommendations contained in this document are those of the authors and are not necessarily endorsed nor should they be interpreted as representing the official policies, either expressed or implied, of the Army Research Office (ARO) or the U.S. Government. The U.S. Government is authorized to reproduce and distribute reprints for Government purposes notwithstanding any copyright notation herein.
\end{acknowledgments}

\appendix

\renewcommand\thefigure{A.\arabic{figure}}
\setcounter{figure}{0}
\setcounter{section}{0}

\section{Derivation of the two-channel effective Hamiltonian} \label{appA}

In the main text, we stated that the effective Hamiltonian $H_{\text{eff}}$ in Eq.~(\ref{HBdGEff}) can be derived by projecting the BdG Hamiltonian $H_{BdG}$ in Eq.~(\ref{HBdG}) onto the two channel orbital wave functions $\chi_L$ and $\chi_R$, which are superpositions of the two lowest-energy orbital subbands $\varphi_1$ and $\varphi_2$ of the effective mass Hamiltonian $H_0$. In this section, we provide details regarding this derivation. For clarity, some statements from the main text will be repeated here.

To begin, we define an orbital subband $\varphi_n$ as the eigenstate of the effective mass Hamiltonian $H_0$ for $k_z = 0$, i.e. 
\begin{equation}
    H_0(0) \varphi_n(x,y) = \varepsilon_n \varphi_n(x,y),
\end{equation}
where $\varepsilon_n$ is the subband energy and $n \in \mathbb{N}^+$. Note that $\varphi_n(x,y)$ can be chosen to be real-valued, which we adopt. Also note that the orbital subband $\varphi_n$ is only an orbital wave function, which does not contain spin. This is well-defined because the effective mass Hamiltonian $H_0$ is spin independent, i.e. $H_0$ is proportional to the identity operator $\sigma_0$ in spin space. We can construct a subband wave function with spin using a simple tensor product, $\varphi_{n,\sigma} = \varphi_n(x,y) \ket{\sigma}$, where $\sigma = \uparrow, \downarrow$. Furthermore, we can define a BdG subband wave function that specifies both the spin and particle-hole degrees of freedom as the tensor product, $\varphi_{n,\sigma,\tau} = \varphi_n(x,y) \ket{\sigma \tau}$, where $\tau = p$ for particle and $\tau = h$ for hole.

We now rewrite our BdG Hamiltonian using the basis consisting of the BdG subband wave functions $\left\{\varphi_{n,\sigma, \tau}\right\}$. This leads to the BdG Hamiltonian,
\begin{equation}
    H^\prime_{BdG}(k_z) = 
     \begin{pmatrix}
    H^\prime_N(k_z) & \widetilde{\Delta}^\prime \\
    -\widetilde{\Delta}^{\prime *} & -H_N^{\prime*}(-k_z)
    \end{pmatrix}, \label{BdGprime}
\end{equation}
where $H_N^\prime(k_z)$ and $\widetilde{\Delta}^\prime$ are matrices whose elements are found by evaluating the expressions, $\mel{\varphi_{m,\sigma}}{H_N(k_z)}{\varphi_{n,\sigma^\prime}}$ and $\mel{\varphi_{m,\sigma}}{\widetilde{\Delta}}{\varphi_{n,\sigma^\prime}}$, respectively. Performing these calculations, we find $H_N^\prime(k_z)$ and $\widetilde{\Delta}^\prime$ can be expressed as
\begin{align}
    H_N^\prime(k_z) &= \bar{\Lambda}(k_z)\sigma_0 + \bar{\alpha}_x k_z \sigma_y - \bar{\alpha}_y k_z \sigma_x, \\
    \widetilde{\Delta}^\prime &= i \bar{\Delta}\sigma_y,
\end{align}
where $\sigma_j$ are Pauli spin matrices. Here, $\bar{\Lambda}, \bar{\alpha}_x$, $\bar{\alpha}_y$, and $\bar{\Delta}$ are real-symmetric matrices with elements,
\begin{align}
    \bar{\Lambda}_{mn}(k_z) &=
    \delta_{mn}\left(\varepsilon_{n} + \frac{\hbar^2 k_z^2}{2m^*}\right), \\
    \left(\bar{\alpha}_j\right)_{mn} &= 
    e C \int \varphi_m(x,y) (\partial_j\phi) \varphi_n(x,y) \,dx dy,  \label{alpMTXelem2} \\
    \bar{\Delta}_{mn} &= \int \varphi_m(x,y) \Delta(x,y) \varphi_n(x,y) \,dx dy.
\end{align}
Notice that $\bar{\Lambda}$ is diagonal in subband space since the kinetic energy term $\hbar^2 k_z^2/(2 m^*)$ in the effective mass Hamiltonian $H_0(k_z)$ has no spatial dependence. The spin-orbit $\bar{\alpha}_j$ and superconductive pairing $\bar{\Delta}$ matrices, in contrast, couple different subbands because of the spatial inhomogeneity of $\phi(x,y)$ and $\Delta(x,y)$. The full BdG Hamiltonian $H_{\text{BdG}}^\prime(k_z)$ can then be expressed as
\begin{equation}
\begin{split}
    H_{\text{BdG}}^\prime(k_z) =& \bar{\Lambda}(k_z)\sigma_0 \tau_z + \bar{\alpha}_x k_z \sigma_y \tau_z - \bar{\alpha}_y k_z \sigma_x \tau_0 \\
    &- \bar{\Delta}\sigma_y \tau_y, 
\end{split}    
    \label{HBdGPrime}
\end{equation}
where $\tau_j$ are Pauli matrices acting in particle-hole space.

The largest energy scale present in Eq.~(\ref{HBdGPrime}) is the energy separation between most subbands, i.e. $\varepsilon_m - \varepsilon_n$ for $m \neq n$. In comparison, the $\bar{\alpha}_j$ and $\bar{\Delta}$ terms that provide inter-subband coupling are small perturbations. We can then arrive at a low-energy effective Hamiltonian that captures all of the relevant low-energy physics by projecting $H_{\text{BdG}}^\prime(k_z)$ onto a few low-energy subbands. As stated in the main text, the two lowest-energy subbands $\varphi_1$ and $\varphi_2$ are superpositions of the left and right channel wave functions $\chi_L$ and $\chi_R$, depicted in Fig. \ref{FIG1}(b) of the main text: $\varphi_n = a_n \chi_L + b_n \chi_R$ for $n = 1,2$. The channel wave functions, being strongly localized near the InAs-Al interfaces as shown in Fig. \ref{FIG1}(b) of the main text, have relatively weak tunnel coupling $t = \mel{\chi_R}{H_0(0)}{\chi_L}$, which leads to a small subband energy difference $\varepsilon_2 - \varepsilon_1 \approx 2t$ between the two lowest-energy subbands. Indeed, $t$ is typically on the same energy scale as $\bar{\alpha}_j k_z$ and $\bar{\Delta}$. We, therefore, need to keep both $\varphi_1$ and $\varphi_2$ in the low-energy basis to capture the relevant low-energy physics. In contrast, $\varepsilon_3 - \varepsilon_2 \gg \varepsilon_2 - \varepsilon_1$ %, (\bar{\alpha}_j)_{mn} k_z, \bar{\Delta}_{mn}$ 
since $\varepsilon_3 - \varepsilon_2 \sim \hbar^2/(m^* \ell^2)$, where $\ell$ is the small length scale associated with the localization of the channel wave functions near each InAs-Al interface. Therefore, the relevant low-energy physics can be fully captured while ignoring subbands $\varphi_n$ for $n \geq 3$. In other words, we only keep the first two rows and columns of the $\bar{\Lambda}, \bar{\alpha}_x$, $\bar{\alpha}_y$, and $\bar{\Delta}$ matrices in Eq.~(\ref{HBdGPrime}) for our low-energy effective Hamiltonian.

%%%%%%%%%%%%%%%%%%%%%%%%%%%%%%%%
%%%%%%%%%%%%%%%%%%%%%%%%%%%%
\begin{figure}[t]
\begin{center}
\includegraphics[width=0.48\textwidth]{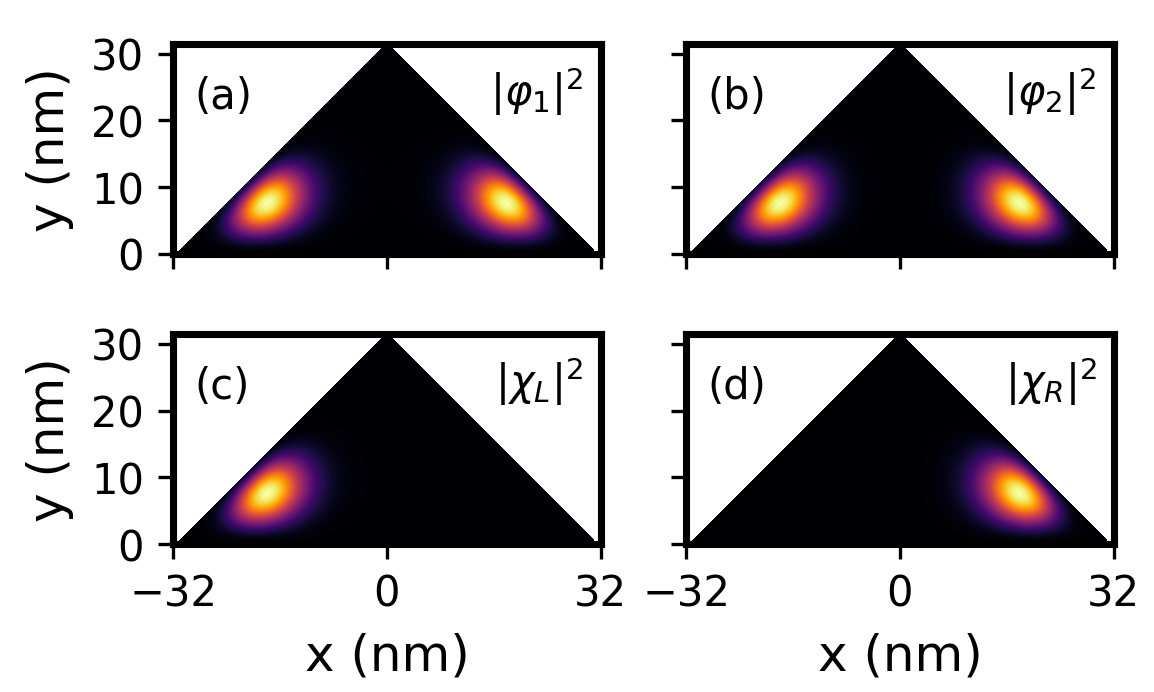}
\end{center}
\vspace{-0.5cm}
\caption{Wave function profiles for the two lowest-energy orbital subbands $\varphi_1$ and $\varphi_2$ are shown in (a) and (b), respectively. These subbands can be decomposed into left and right channels whose wave functions $\chi_1$ and $\chi_2$ are shown in (c) and (d), respectively. Note that these channel wave functions are localized near the InAs-Al interfaces because of the strong band bending near these interfaces, which is primarily due to the band offset between the materials. The gate voltages for this calculation are $V_{T} = -5.5~\text{V}$, $V_{B} = -3~\text{V}$, $V_L = V_R = 0$, and the boundary condition at the InAs-Al interfaces is $\phi = 0.3~\text{V}$. Note that the electrostatic potential $\phi(x,y)$ within the InAs nanowire for these gate settings is shown in Fig.~\ref{FIG1}(c) of the main text.}
\label{FIGA1}
\vspace{-1mm}
\end{figure}
%%%%%%%%%%%%%%%%%%%%%%	
%%%%%%%%%%%%%%%%%%%%%%%%%%%%%%%%

Finally, we transform the low-energy effective Hamiltonian from the basis containing the first two orbital subbands $\varphi_1$ and $\varphi_2$ to the basis containing the two channels $\chi_L$ and $\chi_R$. To do so, we explicitly define the channel wave functions as
\begin{equation}
    \begin{pmatrix}
        \chi_L(x,y) \\ \chi_R(x,y) 
    \end{pmatrix}
        = O
    \begin{pmatrix}
        \varphi_1(x,y)  \\ \varphi_2(x,y) 
    \end{pmatrix},
\end{equation}
where $O$ is an orthogonal matrix. Specifically, we choose $O$ such that the channel wave functions are maximally separated from one another in the $x$ direction, i.e. $\Delta x =  \mel{\chi_R}{x}{\chi_R} - \mel{\chi_L}{x}{\chi_L}$ is maximized. To find such an $O$, we first define $X$ as
\begin{equation}
    X =
    \begin{pmatrix}
        \mel{\varphi_1}{x}{\varphi_1} & \mel{\varphi_1}{x}{\varphi_2} \\
        \mel{\varphi_2}{x}{\varphi_1} & \mel{\varphi_2}{x}{\varphi_2}
    \end{pmatrix},
\end{equation}
i.e. the matrix representation of the $x$ operator within the subspace containing the two lowest-energy orbital subbands. $\Delta x$ is then maximized by choosing $O$ to be the matrix which diagonalizes $X$, $O^T X O = \text{diag}(\widetilde{x}_1, \widetilde{x}_2)$ where $\widetilde{x}_1 < \widetilde{x}_2$. An example is shown in Fig.~\ref{FIGA1} of the wave functions of the two lowest-energy orbital subbands, $\varphi_1$ and $\varphi_2$, and how they decompose into left and right channel wave functions, $\chi_L$ and $\chi_R$. Note that in the case where we have mirror symmetry about the $\hat{y}$ axis ($g=0$), the channel wave functions, $\chi_L$ and $\chi_R$, are simply the even and odd equal superpositions of the orbital subband wave functions, $\varphi_1$ and $\varphi_2$, as is the case in Fig.~\ref{FIGA1}. Evaluating the effective BdG Hamiltonian in this new basis, where the channel wave functions $\chi_L$ and $\chi_R$ replace the orbital subbands $\varphi_1$ and $\varphi_2$, then yields the two-channel effective Hamiltonian,
%\begin{equation}
%    \begin{split}
%    H_{BdG}^\text{eff}(k_z) =&
%        \left[
%        \left(\frac{\hbar^2 k_z^2}{2m^*} - \mu\right)\lambda_0 
%        + t \lambda_x  
%        + g \lambda_z \right]\sigma_0\tau_z \\
%        &- \Delta_0 \sigma_y \lambda_z \tau_y \\
%        &+ \alpha k_z\left( 
%        \cos\theta \sigma_y \lambda_z \tau_z 
%        +\sin\theta \sigma_x \lambda_0 \tau_0 \right)
%         \\
%        &+ \left(\eta_1\lambda_0 + \xi_1\lambda_x \right) k_z \sigma_y\tau_z \\
%        &+ \left(\eta_2\lambda_z + \xi_2\lambda_x \right) k_z \sigma_x  \tau_0 \\
%        %+ \xi_1 k_z \sigma_y \lambda_x \tau_z 
%        %+ \xi_2 k_z \sigma_x \lambda_x \tau_0 
%        &+ (\Delta_1 \lambda_0 + \Delta_2 \lambda_x) \sigma_y \tau_y
%    \end{split}, \label{HBdGEffExtended}
%\end{equation}
\begin{widetext}
\begin{equation}
    \begin{split}
    H_{BdG}^\text{eff}(k_z) =&
        \left[
        \left(\frac{\hbar^2 k_z^2}{2m^*} - \mu\right)\lambda_0 
        + t \lambda_x  
        + g \lambda_z \right]\sigma_0\tau_z
        - \Delta_0 \sigma_y \lambda_z \tau_y
        + \alpha k_z\left( 
        \cos\theta \sigma_y \lambda_z \tau_z 
        +\sin\theta \sigma_x \lambda_0 \tau_0 \right)
         \\
        &+ \left(\eta_1\lambda_0 + \xi_1\lambda_x \right) k_z \sigma_y\tau_z
        + \left(\eta_2\lambda_z + \xi_2\lambda_x \right) k_z \sigma_x  \tau_0 
        %+ \xi_1 k_z \sigma_y \lambda_x \tau_z 
        %+ \xi_2 k_z \sigma_x \lambda_x \tau_0 
        + (\Delta_1 \lambda_0 + \Delta_2 \lambda_x) \sigma_y \tau_y
    \end{split}, \label{HBdGEffExtended}
\end{equation}
\end{widetext}
where $\sigma_j$, $\lambda_j$, and $\tau_j$ with $j = 0,x,y,z$ are Pauli matrices acting in spin, channel, and particle-hole space, respectively. In other words, the first and second columns of a $\lambda_j$ matrix correspond to the left and right channel, respectively. The parameters in Eq.~(\ref{HBdGEffExtended}) are given by the expressions,
\begin{align}
    \mu &= (h_{LL}+h_{RR})/2, \\
    t &= h_{RL}, \\
    g &= (h_{LL}-h_{RR})/2, \\
    \Delta_0 &= (d_{LL}-d_{RR})/2, \\
    \alpha \cos\theta &= (s^{x}_{LL}-s^{x}_{RR})/2, \\
    \alpha \sin\theta &= -(s^{y}_{LL}+s^{y}_{RR})/2,
\end{align}
\begin{align}    
    \eta_1 &= (s^{x}_{LL}+s^{x}_{RR})/2, \\
    \eta_2 &= -(s^{y}_{LL}-s^{y}_{RR})/2, \\
    \xi_1 &= s^{x}_{RL}, \\
    \xi_2 &= -s^{y}_{RL}, \\
    \Delta_1 &= (d_{LL}+d_{RR})/2, \\
    \Delta_2 &= d_{RL},
\end{align}
where $h_{ij} = \mel{\chi_i}{H_0(k_z)}{\chi_j}$, $s^{x}_{ij} = \mel{\chi_i}{\widetilde{\alpha}_x(x,y)}{\chi_j}$, $s^{y}_{ij} = \mel{\chi_i}{\widetilde{\alpha}_y(x,y)}{\chi_j}$, and $d_{ij} = \mel{\chi_i}{\Delta(x,y)}{\chi_j}$ with $i,j = L, R$. Note that Eq.~(\ref{HBdGEffExtended}) contains additional terms involving $\eta_1$, $\eta_2$, $\xi_1$, $\xi_2$, $\Delta_1$, and $\Delta_2$ that are not show in Eq.~(\ref{HBdGEff}) of the main text. Here, $\eta_1$ and $\eta_2$ account for the fact that the $x$ and $y$ components of the electric field at the positions of the two channels are only \textit{approximately} opposite and equal, respectively. The terms $\xi_1$ and $\xi_2$ account for the interchannel spin-orbit coupling, as indicated by the $\lambda_x$ matrix. The $\Delta_1$ and $\Delta_2$ terms allow for differences between the two channels in pairing strength and interchannel pairing, respectively. These six terms in our calculations are all found to be small and do not qualitatively impact the physics of the effective model. For simplicity, we have ignored these terms when showing the effective Hamiltonian in Eq.~(\ref{HBdGEff}) of the main text. However, these terms have been included in the calculation of the topological phase diagrams shown in Figs.~\ref{FIG6}(a) and (b) of the main text. Furthermore, we find that ignoring these terms only slightly changes the phase diagrams with no qualitative differences.

\section{Transformation of the two-channel effective Hamiltonian} \label{TransApp}
In the main text, we stated that the two-channel effective Hamiltonian $H_{\text{eff}}(k_z)$ shown in Eq.~(\ref{HBdGEff}) transforms into $H^\prime_{\text{eff}}(k_z)$ shown in Eq.~(\ref{HeffTrans}). The Hamiltonians are connected by the relation $H^\prime_{\text{eff}}(k_z) = U^\dagger H_{\text{eff}}(k_z) U$, where $U$ is a unitary matrix. Here, we provide the form of $U$ and discuss its structure. The goal of the transformation is to bring the Hamiltonian into a form where the spins are decoupled for the ideal limit of antiparallel spin-orbit vectors ($\theta = 0$).

The unitary matrix $U$ is best understood as the product of four simple transformations, $U = R^{\sigma}_{x} S^{\sigma}_{h} S^{\lambda}_{h} S^{\lambda}_{\downarrow}$, where $R$ and $S$ stand for rotation and swap operations that we now explain. 
%The final form of $U$ is given by 
%\begin{equation}
%    U = \frac{1}{4}(1 + i[\sigma_x + \sigma_y + \sigma_z])(1 - i\sigma_z%lambda_x)(1 + i\lambda_x\tau_z). \label{Uform}
%\end{equation}
%This final form can be understood if we decompose $U$ into a series of simpler transformations. Indeed, let us write the matrix as $U = R^{\sigma}_{x} S^{\sigma}_{h} S^{\lambda}_{h} S^{\lambda}_{\downarrow}$, where $R$ and $S$ stand for rotation and swap operations that we will explain. 
The first operation $R^{\sigma}_{x}$ is a rotation of the spins by $90^\circ$ about the $x$ axis, 
\begin{equation}
    R^{\sigma}_{x} = \frac{1}{\sqrt{2}}\left(1 + i\sigma_x\tau_z\right).
\end{equation}
This results in the transformed Hamiltonian $H_{1}(k_z) =  {R^{\sigma}_{x}}^\dagger H_{\text{eff}}(k_z) R^{\sigma}_{x}$ given by
\begin{equation}
\begin{split}
    H_{1}(k_z) =& \left[
        \varepsilon(k_z)
         \lambda_0  
        + t \lambda_x 
        + g \lambda_z 
        \right]\sigma_0\tau_z 
        - \Delta_0 \sigma_y \lambda_z \tau_y \\
        &+ \alpha k_z \left(\cos\theta \sigma_z \lambda_z \tau_o 
        +  \sin\theta \sigma_x \lambda_0 \tau_0\right),
\end{split}        
\end{equation}
where we see that the spin-orbit coupling aligns with the $z$ axis for the ideal limit of antiparallel spin-orbit vectors ($\theta = 0$). Next, we notice that the superconductive pairing exists between particles and holes of the opposite spin species, as evident in the $\Delta_0 \sigma_y \lambda_z \tau_y$ term. We want, however, the pairing to exist between the same spin species since we are seeking a final Hamiltonian that is block diagonal in spin space for the case of $\theta = 0$. To accomplish this, we swap the spin degrees of freedom in the hole sector using the operator,
\begin{equation}
    S^{\sigma}_{h} = \frac{1}{2}\left[
    \sigma_0\left(\tau_0 + \tau_z\right) 
    + \sigma_x\left(\tau_0 - \tau_z\right)
    \right].
\end{equation}
This results in the transformed Hamiltonian $H_{2}(k_z) =  {S^{\sigma}_{h}}^\dagger H_{1}(k_z) S^{\sigma}_{h}$ given by 
\begin{equation}
\begin{split}
    H_{2}(k_z) =& \left[
        \varepsilon(k_z)
         \lambda_0  
        + t \lambda_x 
        + g \lambda_z 
        \right]\sigma_0\tau_z 
        + \Delta_0 \sigma_z \lambda_z \tau_x \\
        &+ \alpha k_z \left(\cos\theta \sigma_z \lambda_z \tau_z 
        +  \sin\theta \sigma_x \lambda_0 \tau_0\right),
\end{split}        
\end{equation}   
where we see that the Hamiltonian is indeed block diagonal in spin space for the case of $\theta = 0$. Next, we notice that the superconductive pairing within each spin block is diagonal in channel space. We want, however, the channels to play the role of ordinary spins in an s-wave superconductor, which have superconductive pairing between particles and holes of \textit{opposite} spin. In order to make the channel degree of freedom better resemble the ordinary spin degree of freedom, we swap the channel degrees of freedom in the hole sector using the operator,
\begin{equation}
    S^{\lambda}_{h} = \frac{1}{2}\left[
    \lambda_0\left(\tau_0 + \tau_z\right) 
    + \lambda_x\left(\tau_0 - \tau_z\right)
    \right].
\end{equation}
This results in the transformed Hamiltonian $H_{3}(k_z) =  {S^{\lambda}_{h}}^\dagger H_{2}(k_z) S^{\lambda}_{h}$ given by 
\begin{equation}
\begin{split}
    H_{3}(k_z) =& \left[
        \varepsilon(k_z)
         \lambda_0  
        + t \lambda_x 
        \right]\sigma_0\tau_z
        + g \sigma_0 \lambda_z \tau_0
        - \Delta_0 \sigma_z \lambda_y \tau_y \\
        &+ \alpha k_z \left(\cos\theta \sigma_z \lambda_z \tau_o 
        +  \sin\theta \sigma_x \lambda_0 \tau_0\right)
    . \label{H3}
\end{split}    
\end{equation}
Interestingly for $g = 0$, each diagonal spin block of Eq.~(\ref{H3}) is exactly the Lutchyn-Oreg Hamiltonian of Refs.~\cite{Lutchyn2010,Oreg2010} for a superconducting nanowire with spin-orbit coupling in an external magnetic field, except that the spin degrees of freedom are replaced by channels $\sigma_j \rightarrow \lambda_j$, and the Zeeman energy is replaced by the interchannel tunnel coupling  $t$. The diagonal spin blocks are not identical, however, since they have opposite superconductive pairing and spin-orbit coupling coefficients, as evident by the $\sigma_z$ in the $\Delta_0\sigma_z \lambda_y \tau_y$ and $\alpha k_z \cos \theta \sigma_z \lambda_z \tau_o$ terms. The spin blocks become identical if we swap the channel degrees of freedom within the spin $\downarrow$ sector using the operator,
\begin{equation}
    S^{\lambda}_{\downarrow} = 
    \frac{1}{2}\left[
    \left(\sigma_0 + \sigma_z\right) \lambda_0
    + \left(\sigma_0 - \sigma_z\right)\lambda_x
    \right].
\end{equation}
Finally, this results in the effective Hamiltonian $H^\prime_{\text{eff}}(k_z) = {S^{\lambda}_{h}}^\dagger H_{3}(k_z) S^{\lambda}_{h} = U^\dagger H_{\text{eff}}(k_z)U$ shown in Eq.~(\ref{HeffTrans}) of the main text. Notice that this last operation transforms the $g \sigma_0 \lambda_z \tau_o$ term in $H_{3}(k_z)$ into $g \sigma_z \lambda_z \tau_o$, where the channel detuning $g$ is effectively opposite for the two spin blocks. As stressed in the main text, this difference between the spin blocks arising from the channel detuning $g$ is responsible for the increased robustness of the topological phase to spin-orbit vector misalignment ($\theta = 0$). %, as evident in Figs.~\ref{FIG3} and 4 of the main text. 
Lastly, we note that the product of these four simple operations can be algebraically manipulated into the more compact expression,
\begin{equation}
    U = \frac{1}{4}(1 + i[\sigma_x + \sigma_y + \sigma_z])(1 - i\sigma_z\lambda_x)(1 + i\lambda_x\tau_z). \label{Uform}
\end{equation}

\section{Phase diagram calculation method} \label{PDCMapp}

%%%%%%%%%%%%%%%%%%%%%%%%%%%%%%%%
%%%%%%%%%%%%%%%%%%%%%%%%%%%%
\begin{figure}[t]
\begin{center}
\includegraphics[width=0.48\textwidth]{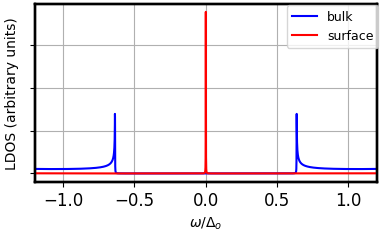}
\end{center}
\vspace{-0.5cm}
\caption{Surface (red) and bulk (blue) local density of states for a system in the topological phase. The presence of Majorana zero modes is evident from the peak in the surface local density of states at zero energy $\omega = 0$. The absence or presence of such a peak can be used to determine if the system is in the trivial or topological phase for any given choice of parameters. System parameters are $m^* = 0.026m_o$, $\Delta_0 = 0.2~\text{meV}$, $t = 0.5~\text{meV}$, $\alpha = 20~\text{meV}\cdot\text{nm}$, and $\theta = 0$. All other parameters in Eq.~(\ref{HBdGEffExtended}) are zero.}
\label{FIGA2}
\vspace{-1mm}
\end{figure}
%%%%%%%%%%%%%%%%%%%%%%	
%%%%%%%%%%%%%%%%%%%%%%%%%%%%%%%%

In the main text, we provide several phase diagrams showing the boundary between the trivial and topological phases, as well as the topological gap $\Delta_{\text{top}}$ within the topological regions. Often times, the assignment of trivial and topological phases is done by calculating a topological invariant~\cite{Chiu2016}. In this work, we use an alternative technique where we look for the presence of zero-energy Majorana zero modes at the end of a semi-infinite system. To do so, we first discretize the Hamiltonian in Eq.~(\ref{HBdGEffExtended}) on a 1D lattice with a small lattice spacing of $a = 0.5~\text{nm}$. Note that each site of the 1D lattice has eight orbitals to account for the spin, channel, and particle-hole degrees of freedom. Information about the local density of states is then contained in the Green's function matrix $G(\omega) = (\omega I - H_{dis} + i \eta)^{-1}$, where $\omega \in \mathbb{R}$ is the input energy, $H_{dis}$ is the discretized Hamiltonian matrix, $I$ is the identity matrix, and $\eta = 10^{-5}~\text{meV}$ is a small energy to bring the Green's function into the upper half of the complex plane. While generic matrix inversion is numerically expensive, finding the diagonal elements of $G(\omega)$ can be done in a numerically efficient manner for systems with translation invariance by using the decimation technique of Ref.~\cite{Sancho1985}. Using this technique, the surface Green's function block $G_s$ (which is just the block of the Green's function matrix for the first lattice site) and the bulk Green's function block $G_b$ can be found for a system with $2^N$ lattice sites using only $N$ decimation iterations. These, in turn, give us the surface and bulk local density of states,
\begin{align}
    LDOS_{s}(\omega) &= -\frac{1}{\pi} \Im( \Tr [G_s(\omega)]), \\
    LDOS_{b}(\omega) &= -\frac{1}{\pi} \Im(\Tr [G_b(\omega)]).
\end{align}
Using $N = 50$ iterations, we can then obtain the surface and bulk density of states for system of length $L \approx 10^6~\text{m}$, which is semi-infinite for all practical purposes. An example of the surface and bulk local density of states for a system in the topological phase is shown in Fig.~\ref{FIGA2}. We see a peak in the surface local density of states at zero energy $\omega = 0$ coming from the presence of Majorana zero modes at the ends of the system. The absence or presence of such a peak can then be used to determine if the system is in the trivial or topological phase for any given choice of parameters. In addition, we can read off the topological gap $\Delta_{\text{top}}$ of the topological phase from the location of the coherence peaks in the bulk local density of states (blue curve in Fig.~\ref{FIGA2}).

\section{Quantifying Dresselhaus and transverse-Rashba spin-orbit coupling} \label{Dressel}

In the main text, we only included Rashba spin-orbit coupling and limited it to the component involving the longitudinal momentum $k_z$ (see Eq.~(3) of main text). However, there also exists Rashba spin-orbit coupling involving the transverse momenta $k_x$ and $k_y$, along with Dresselhaus spin-orbit coupling. Within in the main text, we neglected these terms due to their contributions yielding only small quantitative effects while not affecting the qualitative aspects of the physics. In this section, we quantify these other forms of spin-orbit coupling to show that neglecting them is well justified.

Let us begin with the Dresselhaus spin-orbit coupling, which arises from the bulk inversion asymmetry of the InAs zinc blende lattice. Its Hamiltonian component, which is to be added to $H_{\text{SO}}$ in Eq.~(3) of the main text, takes the form~\cite{Winkler2003},
\begin{equation}
\begin{split}
    H_{\text{BIA}} = \gamma_D \big[&
    \left(k_y^2 - k_z^2\right) k_x \sigma_x
    + \left(k_z^2 - k_x^2\right) k_y \sigma_y \\
    &+ \left(k_x^2 - k_y^2\right) k_z \sigma_z
    big], \label{HBIA}
\end{split}    
\end{equation}
where $\gamma_D$ is a material-dependent constant given by $\gamma_D = 27.18~\text{meV}\cdot\text{nm}^3$ for InAs~\cite{Winkler2003}. (Ref.~\cite{Gmitra2016} gives a similar value of $\gamma_D = 21.7~\text{meV}\cdot\text{nm}^3$.) Note that the Dresselhaus spin-orbit Hamiltonian takes the particular form shown in Eq.~(\ref{HBIA}) because the triangular InAs nanowires studied in this work are grown along the $\left[001\right]$ crystallographic direction with the upper facets of the nanowire being within the $[110]$ and $[\bar{1}10]$ planes~\cite{Wojcik2018}. Otherwise, the appropriate coordinate rotation would have to be applied to Eq.~(\ref{HBIA}). 
%Projecting the Dresselhaus Hamiltonian onto the previously defined two-channel basis and including the particle-hole degree of freedom yields
Projecting the Dresselhaus Hamiltonian onto the two-channel basis defined in Sec. \ref{appA} leads to the term,
\begin{widetext}
\begin{equation}
    H_{\text{BdG}}^{\text{eff,BIA}} = \nu_x \sigma_x \lambda_y \tau_0 + \nu_y \sigma_y \lambda_y \tau_z
    +\left[\left(\frac{\beta_{LL} + \beta_{RR}}{2}\right)\lambda_0 + \left(\frac{\beta_{LL} - \beta_{RR}}{2}\right)\lambda_z
    + \beta_{LR}\lambda_x
    \right]k_z \sigma_z  \tau_0
    + \mathcal{O}(k_z^2),  
\end{equation}
\end{widetext}
which should be added to $H_{\text{BdG}}^{\text{eff}}$ in Eq.~(\ref{HBdGEffExtended}). Here, $\nu_x$ and $\nu_y$ are interchannel spin-orbit coupling coefficients that come from the two terms in Eq.~(\ref{HBIA}) not involving the longitudinal momentum $k_z$ and are given by
\begin{align}
    \nu_x &= -i\gamma_D\mel{\chi_R}{k_x k_y^2}{\chi_L}, \\
    \nu_y &= i\gamma_D\mel{\chi_R}{ k_x^2 k_y}{\chi_L},
\end{align}
where $\chi_L$ and $\chi_R$ are the channel wave functions. Note that these these terms do not produce analogous \textit{intrachannel} spin-orbit coupling coefficients. This is because the expectation values of the operators $k_x k_y^2$ and $k_x^2 k_y$ automatically vanish since they both have an imaginary prefactor when expressed as differential operators and the channel wave functions $\chi_i$ are real-valued. %Note that only interchannel coupling is allowed for the terms involving $\sigma_x$ and $\sigma_y$ in Eq.~(\ref{HBIA}) because the expectation values of the operators $k_x k_y^2$, $k_x^2 k_y$, $k_x$, and $k_y$ all vanish since the channel wave functions $\chi_i$ are real-valued. 
The $\beta_{ij}$ coefficients come from the term involving $\sigma_z$ in Eq.~(\ref{HBIA}) and are given by
\begin{equation}
    \beta_{ij} = \gamma_D \mel{\chi_i}{k_x^2 - k_y^2}{\chi_j}. \label{betaij}
\end{equation}
Note that both interchannel $\beta_{LR}$ and intrachannel $\beta_{ii}$ coefficients exist since the expectation value of the operator $k_x^2 - k_y^2$ for each channel wave function generically does not vanish.

%%%%%%%%%%%%%%%%%%%%%%%%%%%%%%%%
%%%%%%%%%%%%%%%%%%%%%%%%%%%%
\begin{figure}[t]
\begin{center}
\includegraphics[width=.48\textwidth]{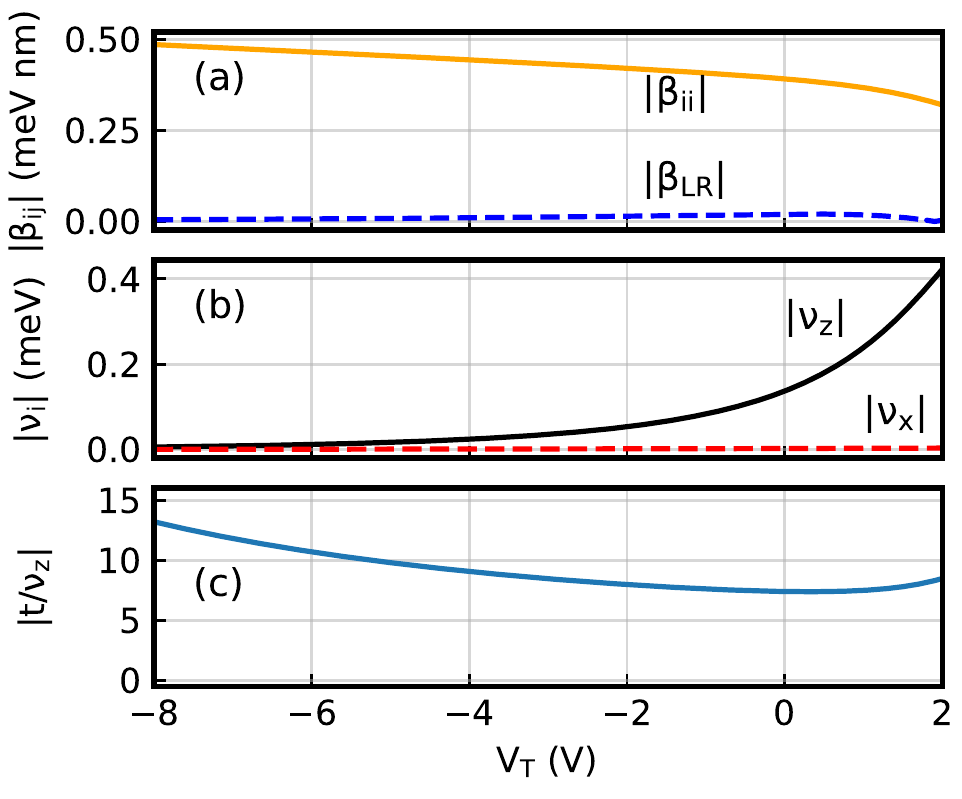}
\end{center}
\vspace{-0.5cm}
\caption{Spin-orbit coefficients from Dresselhaus and transverse-Rashba spin-orbit coupling as a function of topgate voltage $V_{T}$ for $\mu = g = 0$. %The backgate voltage $V_B$ for any given value of $V_T$ is used to tune the chemical potential to $\mu = 0$, which corresponds to having one occupied subband with the subband energies for $k_z = 0$ being $\varepsilon_n = \pm \sqrt{t^2 + g^2}$ for the two lowest-energy subbands, $n = 1,2$. The side gates are set to $V_L = V_R = 0$. 
(a) Intrachannel $\beta_{ii}$ (solid gold) and interchannel $\beta_{LR}$ Dresselhaus coefficients (dashed blue). The intrachannel Dresselhaus coefficient $\beta_{ii}$ dominates for the interchannel Dresselhaus coefficient $\beta_{LR}$. Furthermore, the intrachannel Dresselhaus coefficient $\beta_{ii}$ is smaller than the numerically extracted Rashba spin-orbit coefficient of $\alpha\approx 25~\text{meV}\cdot\text{nm}$ (given in the main text) by over a factor of 40. (b) Interchannel spin-orbit coefficients $\nu_x$ and $\nu_z$ that come from Dresselhaus and transverse-Rashba spin-orbit coupling, respectively. (c) Ratio of interchannel tunnel coupling $t$ and interchannel spin-orbit coefficient $\nu_z$, showing that the spin-independent coupling $t$ dominates over the spin-dependent coupling $\nu_z$.}
\label{FIGA3}
\vspace{-1mm}
\end{figure}
%%%%%%%%%%%%%%%%%%%%%%	
%%%%%%%%%%%%%%%%%%%%%%%%%%%%%%%%

The values of the $\beta_{ij}$ coefficients are shown in Fig.~\ref{FIGA3}(a) as a function of topgate voltage $V_{T}$ for $\mu = g = 0$. %Similar to Fig.~\ref{FIG6} of the main text, the backgate voltage $V_B$ for any given value of $V_T$ is used to tune the chemical potential. The various spin-orbit coefficients are extracted for $\mu = 0$, which corresponds to having one occupied subband with the subband energies for $k_z = 0$ being $\varepsilon_n = \pm \sqrt{t^2 + g^2}$ for the two lowest-energy subbands, $n = 1,2$. The side gates are set to $V_L = V_R = 0$.
The interchannel coefficient $\beta_{LR}$ and intrachannel coefficient $\beta_{ii}$ are shown as blue (dashed) and gold (solid) lines, respectively. Note that $\beta_{LL} = \beta_{RR}$ because of mirror symmetry about the $\hat{y}$ axis when $g = 0$. Notice that the interchannel coefficient $\beta_{LR}$ is smaller than the intrachannel coefficient $\beta_{ii}$ by at least an order of magnitude over the entire voltage range. Furthermore, comparing the intrachannel Dresselhaus coefficient $\beta_{ii}$ to the extracted Rashba spin-orbit coefficient of $\alpha \approx 25~\text{meV}\cdot\text{nm}$ from the main text, we see that $\beta_{ii}$ is smaller by over a factor of $40$. Therefore, we conclude that the Dresselhaus terms involving $\beta_{ij}$ %and the longitudinal momentum $k_z$ 
have a negligible effect on the physics. In addition, the interchannel spin-orbit coefficient $\nu_x$ is shown in Fig.~\ref{FIGA3}(b) as a function of $V_{T}$, as a red (dashed) line. We numerically find that $\nu_x$ is always small ($\nu_x \lesssim 1~\mu\text{eV}$), and is therefore safely neglected. Finally, we also numerically found that $\nu_y$ from Dresselhaus spin-orbit coupling vanishes, and, therefore, plays no role in our system.

It is interesting to note that the physical reason why the intrachannel coefficients $\beta_{ii}$ are so small in this system is because the upper facets of the InAs nanowire are within the $[110]$ and $[\bar{1}10]$ planes. To understand this, notice first that the left channel wave function $\chi_L$ in Fig.~\ref{FIGA1}(c) closely resembles a particle-in-a-box state with quantization axes that are $\approx 45^\circ$ with respect to the $x$ and $y$ axes, respectively, that is, along the $[110]$ and $[\bar{1}10]$ crystallographic directions. Next, we can rewrite the $k_x^2 - k_y^2$ operator appearing in Eq.~(\ref{betaij}) as 
\begin{equation}
    k_x^2 - k_y^2 = 2 k_{x^\prime} k_{y^\prime},
\end{equation}
where the $x^\prime$ and $y^\prime$ axes align with the $[110]$ and $[\bar{1}10]$ crystallographic directions. Importantly, a particle-in-a-box state with quantization axes along the $x^\prime$ and $y^\prime$ axes has vanishing expectation value for the operator $k_{x^\prime} k_{y^\prime}$. This is because the the operators $k_{x^\prime}$ and $k_{y^\prime}$ have imaginary prefactors when expressed as differential operators, and the particle-in-a-box state decomposes into a product of real wave functions in the $x^\prime$ and $y^\prime$ direction. Now, $\chi_L$ is not perfectly captured by a product state separable in the $x^\prime$ and $y^\prime$ direction, so $\beta_{LL}$ does not exactly vanish. Its close resemblance, however, keeps $\beta_{LL}$ very small.

Next, let us consider the transverse-Rashba spin-orbit coupling. The Rashba spin-orbit coupling included within $H_{\text{SO}}$ in Eq.~(3) of the main text only involves longitudinal momentum $k_z$, whereas transverse-Rashba spin-orbit coupling involves the transverse momenta $k_x$ and $k_y$. Its Hamiltonian component, which is to be added to $H_{\text{SO}}$ in Eq.~(3) of the main text, is given by~\cite{Escribano2020}
\begin{equation}
\begin{split}
    H_{R_{\perp}} = 
    \frac{1}{2} \Big(&
    \left[\alpha_y(x,y) k_x + k_x \alpha_y(x,y)\right] \\
    &-\left[\alpha_x(x,y) k_y + k_y \alpha_x(x,y)\right]
    \Big)\sigma_z,
\end{split}    
\end{equation}
where care has been taken to properly symmetrize the operators such that the Hamiltonian is Hermitean. %Like $H_{\text{BIA}}$, $H_{R,\text{trans}}$ should be understood as being added to the normal Hamiltonian $H_N$ in the BdG Hamiltonian of Eq.~(\ref{HBdG}) of the main text.
Projecting the transverse-Rashba Hamiltonian onto the two-channel basis defined in Sec. \ref{appA} leads to the term,
\begin{equation}
    H_{\text{BdG}}^{\text{eff},R_{\perp}} = \nu_z \sigma_z \lambda_y \tau_0,
\end{equation}
which should be added to $H_{\text{BdG}}^{\text{eff}}$ in Eq.~(\ref{HBdGEffExtended}). Here, $\nu_z$ is an interchannel spin-orbit coupling coefficient given by
\begin{equation}
\begin{split}
    \nu_z =& \frac{i}{2} \mel{\chi_R}{ \left(\alpha_y(x,y) k_x + k_x \alpha_y(x,y)\right)}{\chi_L}  \\
    &-\frac{i}{2} \mel{\chi_R}{\left(\alpha_x(x,y) k_y + k_y \alpha_x(x,y)\right)}
    {\chi_L}.
\end{split}    
\end{equation}
Note that the transverse-Rashba Hamiltonian does not produce an analogous \textit{intrachannel} spin-orbit coupling coefficient. This is because the expectation values of the operators $k_x$ and $k_y$ automatically vanish for the same reason given above for the $k_x k_y^2$ and $k_x^2 k_y$ operators.

%The values of the $\beta_{ij}$ coefficients involved in Dresselhaus spin-orbit coupling within the one-subband occupancy region of Fig.~\ref{FIG4} (a) of the main text are shown in Fig.~\ref{FIGA3} (a) as a function of topgate voltage $V_{T}$. The back-gate voltage $V_{BG}$ for any value of $V_{T}$ is given by the black dashed line in Fig.~\ref{FIG4} (a) of the main text. 

%%%%%%%%%%%%%%%%%%%%%%%%%%%%%%%%
%%%%%%%%%%%%%%%%%%%%%%%%%%%%
\begin{figure}[t]
\begin{center}
\includegraphics[width=.48\textwidth]{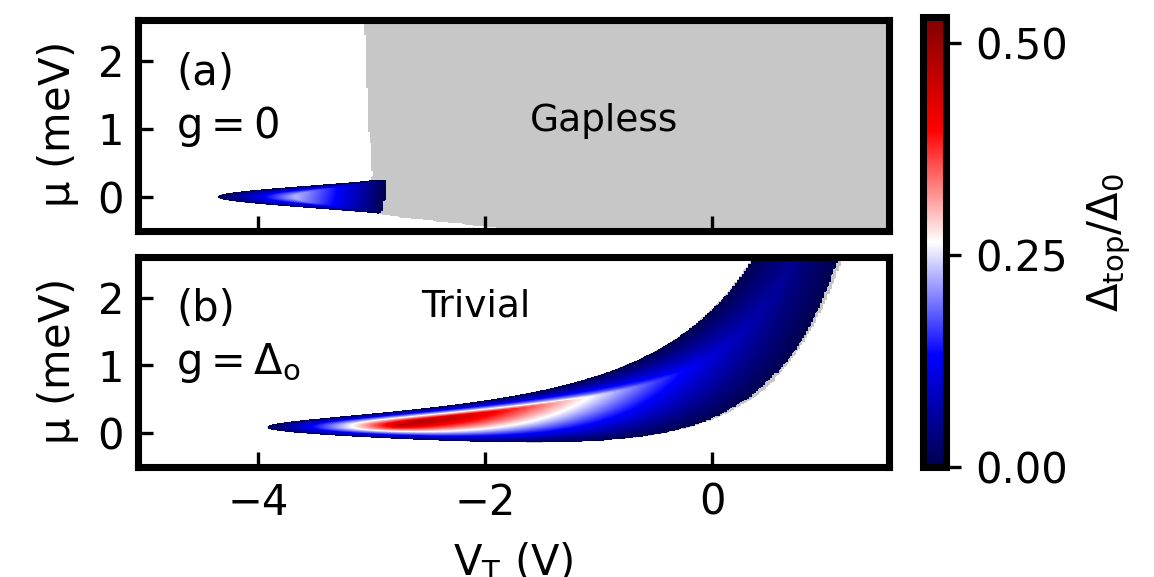}
\end{center}
\vspace{-0.5cm}
\caption{Topological phase diagram of the same system as Fig.~\ref{FIG6} of the main text except that the $\nu_z \sigma_z \lambda_y \tau_0$ term from transverse-Rashba spin-orbit coupling is included in the model. Comparing the results to Figs.~\ref{FIG6}(a) and (b) of the main text, we find that inclusion of the transverse-Rashba spin-orbit coupling slightly reduces the area of the topological phase and the topological gap $\Delta_{\text{top}}$ within the topological region. The effect is quite modest although.}
\label{FIGA4}
\vspace{-1mm}
\end{figure}
%%%%%%%%%%%%%%%%%%%%%%	
%%%%%%%%%%%%%%%%%%%%%%%%%%%%%%%%

The interchannel spin-orbit coefficient $\nu_z$ is shown in Fig.~\ref{FIGA3}(b) as a function of $V_{T}$, as a black (solid) line. We find that the interchannel spin-orbit coefficient $\nu_z$ coming from transverse-Rashba spin-orbit coupling is comparable to the superconducting gap $\Delta_0 = 0.2~\text{meV}$ for the larger top gate voltages in Fig.~\ref{FIGA3}(b). It may then at first seem we cannot ignore the transverse-Rashba spin-orbit coupling. However, if we compare this coupling to the interchannel coupling $t$, as shown in Fig.~\ref{FIGA3}(c), we see that $\nu_z$ is smaller by roughly an order of magnitude within the entire voltage range. We therefore do not expect $\nu_z$ to play a significant role in the physics of the system. To test this expectation, we redo the calculation of the topological phase diagrams shown in Figs.~\ref{FIG6}(a) and (b) of the main text with the inclusion of the $\nu_z \sigma_z \lambda_y \tau_0$ term into the effective two-channel Hamiltonian given in Eq.~(\ref{HBdGEffExtended}). The results are shown in Fig.~\ref{FIGA4}, again for two values of channel detuning, (a) $g = 0$ and (b) $g = \Delta_0$. Comparing the results to Figs.~\ref{FIG6}(a) and (b) of the main text, we see that the addition of transverse-Rashba spin-orbit coupling has a modest effect on the topological phase diagrams. The results are qualitatively the same as the results in Figs.~\ref{FIG6}(a) and (b) of the main text, with the only quantitative differences being a small reduction of the topological region and the reduction of the topological gap $\Delta_{\text{top}}$ within the topological region. Therefore, we conclude that while transverse-Rashba spin-orbit coupling is slightly detrimental to topological superconductivity in our system, the effect is quite small and does not qualitatively impact the physics.

\end{document}